\documentclass[12pt]{article}
\usepackage{epsf,array}
\begin{document}
\begin{titlepage}
\rightline{QMW-PH-97-13}
\rightline{NI-97023}
\def\today{\ifcase\month\or
        January\or February\or March\or April\or May\or June\or
        July\or August\or September\or October\or November\or December\fi,
  \number\year}
%\rightline{\today}
\rightline{hep-th/9705011}
\vskip 1cm
\centerline{\Large \bf  Intersecting Branes
}
\vskip 0.8cm
\centerline{\sc Jerome P. Gauntlett}

\centerline {{\it Physics Department}}
\centerline{{\it Queen Mary and Westfield College}}
\centerline{{\it Mile End Rd, London E1 4NS, U.K.}}
\centerline{\it and}
\centerline {{\it Isaac Newton Institute}}
\centerline{{\it 20 Clarkson Rd}}
\centerline{{\it Cambridge, CB3 0EH, U.K.}}
\centerline{{j.p.gauntlett@qmw.ac.uk }}

\centerline{\sc Abstract}
BPS configurations of intersecting branes have many
applications 
in string theory. We attempt to provide an introductory
and
pedagogical review 
of supergravity solutions corresponding
to orthogonal BPS intersections of branes with an emphasis on eleven
and ten space-time dimensions. Recent work on BPS solutions corresponding
to non-orthogonally intersecting branes is also discussed. 
These notes are based on lectures given at the APCTP
Winter School ``Dualities of Gauge and String Theories",
Korea, February 1997.
\end{titlepage}
%%%%%%%%%%%%%%%%%%%%%%%%%%%%%%%%%%%%%%%%%%%%%%%%%%%%%%%%%%%%%%%%%%
\newpage

\def\beq{\begin{equation}}
\def\eeq{\end{equation}}
\def\bea{\begin{eqnarray}}
\def\eea{\end{eqnarray}}
\renewcommand{\arraystretch}{1.5}
\def\ba{\begin{array}}
\def\ea{\end{array}}
\def\bce{\begin{center}}
\def\ece{\end{center}}
\def\nn{\noindent}
\def\nonu{\nonumber}
\def\pbx{\partial_x}

\font\mybb=msbm10 at 12pt
\def\bb#1{\hbox{\mybb#1}}
\def\bZ {\bb{Z}}
\def\bR {\bb{R}}
\def\bE {\bb{E}}
\def\bT {\bb{T}}
\def\bM {\bb{M}}
\def\bH {\bb{H}}
\def\hk {hyper-K{\" a}hler}
\def\HK {Hyper-K{\" a}hler}
\def\bfomeg{\omega\kern-7.0pt\omega}
\def\bfOmeg{\Omega\kern-8.0pt\Omega}

\def \Q {{\cal Q}}
\def \H {\td H}
\def \la{\longrightarrow}
\def \up {\uparrow}
 \def \upa {\uparrow}
 \def \nea {\nearrow}
\def \pa {\Vert}
\def\ma{\mapsto}
\def\inv{^{-1} }
\def \K {{\tilde K}}
\def\ww {\omega _{km } }
\def\mn {\mu\nu}

\def\np {{  Nucl. Phys. }}
\def\NP {{  Nucl. Phys. }}
\def \pl {{  Phys. Lett. }}
\def \PL {{  Phys. Lett. }}
\def \mpl {{ Mod. Phys. Lett. }}
\def \prl {{  Phys. Rev. Lett. }}
\def \PRL {{  Phys. Rev. Lett. }}
\def \pr  {{ Phys. Rev. }}
\def \PR  {{ Phys. Rev. }}
\def \ap  {{ Ann. Phys. }}
\def \cmp {{ Commun.Math.Phys. }}
\def \CMP {{ Commun.Math.Phys. }}
\def \ijmp {{ Int. J. Mod. Phys. }}
\def \jmp {{ J. Math. Phys.}}
\def \cqg {{ Class. Quant. Grav. }}
\def \CQG {{ Class. Quant. Grav. }}

%Greek Symbols
% GREEK

\def\p{\partial}
\def\del{\partial }
\def\a {\alpha}
\def\b{\beta}
\def\g{\gamma} 
\def\Ga{\Gamma}
\def\de{\delta} \def\De{\Delta}
\def\ep{\epsilon}
\def\vep{\varepsilon}
\def\ze{\zeta}
\def\et{\eta}
\def\t{\theta} \def\Th{\Theta}
\def\vth{\vartheta}
\def\io{\iota}
\def\ka{\kappa}
\def\l{\lambda} 
\def\La{\Lambda}
\def\rh{\rho}
\def\s{\sigma} \def\Si{\Sigma}
\def\ta{\tau}
\def\up{\upsilon} 
\def\Up{\Upsilon}
\def\Ph{\Phi}
\def\vph{\varphi}
\def\ch{\chi}
\def\ps{\psi} 
\def\Ps{\Psi}
\def\om{\omega} 
\def\Om{\Omega}

\def \la{\longrightarrow}
\def\ha { {\textstyle {1\over 2}} }

\def\vol#1{{\bf #1}}
\def\nupha#1{Nucl. Phys., \vol{#1} }
 \def\CMP#1{Comm. Math. Phys., \vol{#1} }
\def\phlta#1{Phys. Lett., \vol{#1} }
\def\phyrv#1{Phys. Rev., \vol{#1} }
\def\PRL#1{Phys. Rev. Lett., \vol{#1} }
\def\prs#1{Proc. Roc. Soc., \vol{#1} }
\def\PTP#1{Prog. Theo. Phys., \vol{#1} }
\def\SJNP#1{Sov. J. Nucl. Phys., \vol{#1} }
\def\TMP#1{Theor. Math. Phys., \vol{#1} }
\def\ANNPHY#1{Annals of Phys., \vol{#1} }
\def\PNAS#1{Proc. Natl. Acad. Sci. USA, \vol{#1} }

\def\mapa#1{\smash{\mathop{\longrightarrow }\limits_{#1} }}
\def\x {\times}

%%%%%%%%%%%%%%%%%%%%%%%
\section{Introduction}
%%%%%%%%%%%%%%%%%%%%%%%%

There is now very strong evidence that an eleven dimensional
$M$-theory plays a fundamental role in string theory (see \cite{townlect}
for a recent review).
%the strongly coupled
%ten-dimensional
%type IIA string theory is an eleven
%dimensional theory called $M$-theory.
The low-energy
limit of $M$-theory is $D$=11
supergravity but it is not yet known what the correct
underlying microscopic theory is\footnote{It has been proposed that $M$-theory
in the infinite momentum frame 
is given by 
the large $N$ limit of a certain quantum mechanics based
on $N\times N$ matrices \cite{suss}.
This interesting
development was 
discussed by H. Verlinde in his lectures at the School and we refer
the reader to his article for more details.}.
It is known that $D$=11 supergravity and hence 
$M$-theory 
contains
solitonic membranes,
``$M2$-branes", and fivebranes, ``$M5$-branes", which play an important
role in the dynamics of the theory. Both of these solitons preserve 
1/2 of the supersymmetry and hence are BPS states.
BPS states are states that preserve some supersymmetry and
are an important 
class of states as we have some control over their
behaviour as various moduli are allowed to vary. It is an
important issue to understand the spectrum of BPS states
in $M$-theory and we will see that there is a large class of states
corresponding to intersecting $M2$-branes and $M5$-branes.

String theory in $D$=10 also contains a rich spectrum of BPS
branes. 
In the type IIA and IIB theories there are branes that carry charges
arising from both the Neveu-Schwarz-Neveu-Schwarz ($NSNS$)
and the Ramond-Ramond ($RR$) sectors of the world-sheet theory.
The former class consists of the fundamental strings and the
solitonic fivebranes, ``$NS5$-branes".
The second class of branes, the ``$D$-branes", have
a simple perturbative description as
surfaces in flat space where open strings can end,
which has played a central role in recent string theory
developments
\cite{polch}. 
By dimensionally reducing the intersecting brane solutions of 
$M$-theory we obtain type IIA solutions corresponding to
intersecting $NS$- and $D$-branes.
Various string dualities then enable one 
to construct
all of the supergravity solutions corresponding to intersecting
branes in both the type IIA and IIB theories. 
The properties of these supergravity solutions
complement what we can learn about the various branes using 
string perturbation theory.

Since solitons are a key ingredient in duality studies, 
a great deal of effort has been devoted to constructing 
general soliton solutions of supergravity theories in various 
dimensions. The intersecting brane solutions in $D$=11 and $D$=10
provide a unified viewpoint since many of the other 
soliton solutions can be obtained by dimensional reduction and
duality transformations. Understanding the general structure of
intersecting brane solutions is an involved task:
a partial list of references is:
%\cite{paptown}-
%,tseytlinharm,klebtseyt,gkt,bbj,kklp,glpt,cvetictseyt,lupope,
%lupopetoda,bderoopanda,paptownkk,costacomp,avv,costablack,av,russotseyt,
%tseytlincomp,zmlz,llp,bergmult,im,costapap,cveticrot,ar,os,englert,aiv,air,
%tseytlincomptwo,ohta,bc,ggpt,
\cite{paptown}-\cite{ballarsen}.
In these lectures we will only consider BPS intersections, but we 
note here that non-BPS solutions have also been studied. 
BPS intersecting branes fall into two categories which have been
termed ``marginal" and ``non-marginal" \cite{tseytlincomptwo}.
Roughly speaking the mass (or tension) $M$
and charges $Q_i$ of marginal configurations satisfy $M=\Sigma Q_i$
while 
for the non-marginal cases one has 
$M^2=\Sigma Q_i^2$, corresponding to non-zero binding
energy.
For the most part 
we will be focusing on the marginal intersections.
The non-marginal solutions can be
obtained from the marginal solutions by dimensional
reduction and/or duality transformations. They were discussed by
J. Russo in his lectures at this school.

Our main focus will be on supergravity solutions
with an emphasis on $M$-theory. 
It is worth pointing out in advance that 
there are a number of important applications 
of intersecting brane configurations which we will not be
discussing in much detail. Let us briefly highlight just two here.
The first
is to provide a microscopic state counting interpretation
of black hole entropy\cite{stromvafa}\footnote{See S. Das's contribution to
the proceedings for more details and references.}.
One can construct classical solutions corresponding to intersecting
$D$-branes that give rise upon dimensional reduction to black holes
with non-zero Bekenstein-Hawking entropy. By exploiting the
perturbative $D$-brane point of view one can count the number of open string
microstates that give rise to the same macroscopic quantum numbers
that the black hole carries and one finds perfect agreement.
It should be noted that while
the perturbative calculation is valid at weak coupling
the supergravity black hole spacetime is valid at strong coupling
and one must invoke 
supersymmetry to argue that the state counting calculation is unchanged
as one varies the coupling. Although this is an exciting development
there is still more to be understood on how these
two complementary views of black holes are related.

A second application is to use BPS intersecting branes to study the
infrared dynamics of supersymmetric gauge theories
\cite{HW,egkrs} (and references therein). 
One considers
different types of branes intersecting in an appropriately chosen
arrangement.
The low-energy dynamics on the world-volume of one type of
brane is associated with a supersymmetric
quantum field theory that one wishes to study.
By considering the low-energy dynamics from the point
of view of different branes and allowing
the branes to move around, enables one, in 
certain cases, to determine the low-energy effective
dynamics of the field theory.
This has proven to be a very powerful tool to study supersymmetric
gauge theories in three and four spacetime dimensions. It is worth
noting that in a 
recent development some aspects of
the supergravity solutions of branes in
$M$-theory played an important  role \cite{wittenmth}.

The plan of the rest of the paper is as follows.
In section 2 we discuss orthogonal 
intersections of branes in $M$-theory. In section 3 
we discuss the intersections of $NS$- and $D$-branes 
in type IIA and IIB string theory. Section 4 reviews
recent solutions on supersymmetric configurations of
branes that intersect non-orthogonally and section 5
concludes.

%%%%%%%%%%%%%%%%%%%%%%%
\section{Intersecting $M$-Branes}
%%%%%%%%%%%%%%%%%%%%%%%%
\subsection{$M2$-branes and $M5$-branes}

The low-energy effective action of $M$-theory is $D$=11
supergravity. 
The bosonic field content consists of a metric, $g_{MN}$,
and a three-form potential, $A_{MNP}$, with four-form field
strength
$F_{MNPQ}=24\nabla_{[M}{A}_{NPQ]}$.
The action for the bosonic fields is given by
\beq
S=\int\sqrt{-{g}}\left\{
R-{1\over 12}{F}^2-{1\over 432}
\epsilon^{M_1\dots M_{11}}{F}_{M_1\dots M_4}
{F}_{M_5\dots M_8}
{A}_{M_9\dots M_{11}}\right\}.
\label{eq:one}
\eeq
Supersymmetric solutions to the corresponding equations of motion
can be constructed
by looking for bosonic backgrounds that admit Killing spinors i.e.,
backgrounds which admit 32-component Majorana spinors $\epsilon$
such that the supersymmetry
variation of the gravitino field $\psi_M$ vanishes:
\beq
\left[D_M+
{1\over 144}({\Gamma_M}^{NPQR}-8\delta^N_M\Gamma^{PQR})
F_{NPQR}\right]\epsilon
=0.
\label{eq:two}
\eeq

The $M2$-brane solution \cite{duste} takes the form
\bea
ds^2&=&H^{1/3}[H^{-1}\left(-dt^2+dx_1^2+dx_2^2\right)
+\left(dx_3^2+\dots dx_{10}^2\right)]\nonu\\
F_{t12\alpha}&=&{c\over 2}{\partial_\alpha H\over H^2},\qquad
H= H(x_3,\dots ,x_{10}), \qquad \nabla^2H=0,\qquad c=\pm1.
\label{eq:three}
\eea
We have 
written the metric with an overall conformal factor as this
form will be convenient when we discuss intersecting $M$-branes.
The solution admits Killing spinors of the form $\epsilon=H^{-1/6}\eta$
with the constant spinor $\eta$ satisfying 
\beq
\hat\Gamma_{012}\eta=c\eta,
\eeq
where $\hat\Gamma_{0\dots p}\equiv\hat\Gamma_0\dots\hat\Gamma_p$
is the product
of $p+1$ distinct Gamma matrices in an orthonormal frame.
Using the fact that $(\hat\Gamma_{012})^2=1$ and that 
${\rm Tr}\hat\Gamma_{012}=0$ we conclude that the $M2$-brane solution has
16 Killing spinors and preserves (breaks) half of the supersymmetry.
The solution is governed by a single
harmonic function that depends on the coordinates 
$\vec x=\{x_3,\dots,x_{10}\}$ and we first take it to be of the form
\beq
H = 1 + {a\over r^6},\qquad
r= \left |\vec{x}\right |. 
\eeq
The solution 
then describes a single $M2$-brane with world-volume oriented along
the $\{0,1,2\}$ hyperplane located at $r=0$. 
The $M2$-brane carries 
electric four-form charge $Q_e$ 
which is defined as the integral of the seven-form\footnote{
To be more precise we should integrate $*F+A\wedge F$, since
the field equation is $d*F+F\wedge F=0$.}
$*F$ around a seven-sphere that surrounds the brane and is 
proportional to $ca$.
If $c=1$ we have
an $M2$-brane, while if $c=-1$ we have an anti-$M2$-brane.
We will
often not distinguish between branes and antibranes in the
following. 
The ADM mass per unit area
or ADM tension $T$ can be calculated and is proportional to 
$|Q_e|$ as one requires for
a BPS state. 
The metric appears to be singular at $r=0$. However, it has been shown
that this surface is in fact a regular degenerate event horizon \cite{dgt}.
The metric can be
continued into an interior region and it is here that a real 
curvature singularity is located.
By generalising the harmonic function to have many centres
\beq
H = 1 + \sum_{I=1}^{k}{a_{I}\over r_{I}^4},\qquad
r_{I}= \left |\vec{x}-\vec{x}_{I}\right |,
\eeq
we obtain $k$ parallel $M2$-branes located at positions 
$\vec{x}_{I}$.

The construction of the $M5$-brane solution \cite{guven}
runs along similar lines.
The solution is given by
\bea
ds^2&=&H^{2/3}\left[
H^{-1}\left(-dt^2 +dx_1^2 +\dots dx_5^2\right)+
\left(dx_6^2+\dots + dx_{10}^2\right)\right]\nonu\\
{F}_{\a_1\dots\a_4}&=&{c\over 2}\ep_{\a_1\dots\a_5}\p_{\a_5}H,
\qquad H=H_i(x_6,\dots,x_{10}),\qquad c=\pm 1,
\eea
where $\epsilon_{\a_1\dots\a_5}$ is the flat $D$=5 alternating symbol.
It again admits 16 Killing spinors given by $\epsilon
=H^{-1/12}\eta$ where $\eta$ now satisfies the projection: 
\beq
\hat\Gamma_{012345}
\eta=c\eta.
\eeq
For a single $M5$-brane we choose the harmonic functions to be
\beq
H = 1 + {a\over r^4},\qquad
r= \left |\vec{x}\right |, 
\eeq
where $\vec{x}= \{x_6,\dots,x_{10}\}$.
The $M5$-brane carries magnetic four-form charge $Q_m$ which is obtained by
integrating $F$ around a four-sphere that surrounds the $M5$-brane and is
proportional to $ca$. $c=\pm 1$ correspond to an $M5$- and an anti-$M5$-brane
respectively.
The ADM tension is again proportional to $|Q_m|$ 
in line with unbroken supersymmetry.
The $M5$-brane is a completely regular solution as was shown in
\cite{ght}.
A configuration of parallel multi-$M5$-branes is obtained by
generalising the single centre harmonic function to have many
centres.

The dimensional reduction of $D$=11 supergravity on a circle
leads to $D$=10 type IIA supergravity. Indeed this is necessary for 
the type IIA string theory to be dual to $M$-theory.
There are two distinct ways in which the $M$-brane solutions
can be dimensionally reduced to $D$=10: they can be ``wrapped"
or ``reduced", as we now explain (we will also return to this
in section 3).
Since both the $M2$-brane and the $M5$-brane solutions
are independent of the coordinates tangent to the world-volume of the branes
we can demand that one of them is a periodic spatial coordinate upon
which we compactify. The result of this wrapping leads to the fundamental
string and
the $D4$-brane solutions of the type IIA theory, respectively.
If we denote the compactified
direction as $x_{10}$ and the other coordinates by $x_\mu$, 
we find that the membrane carries electric two-form
$A_{\mu\nu 10}$ charge while the four-brane carries magnetic three-form
$A_{\mu\nu\rho}$
charge.
The process of reducing along
a direction transverse to the world-volume is slightly more involved.
To obtain a solution that is periodic in such a direction, $x_{10}$ say,
we construct a periodic array of either 
$M2$- or $M5$-branes i.e., we take a multi $M$-brane solution
with the branes lined up along the $x_{10}$
direction and equally spaced by a distance $2\pi R$.
The solution obtained by dimensional reduction along the $x_{10}$ direction
will have non-trivial
dependence on the
compactified coordinate or equivalently the $D$=10 solution will 
have massive Kaluza-Klein modes excited. If we average over
the compact coordinate, i.e., if we ignore the massive  modes, then we
obtain the $D2$-brane and the $NS5$-brane solutions
of type IIA
supergravity, respectively. The former carries 
electric $A_{\mu\nu\rho}$ charge and the latter
magnetic $A_{\mu\nu 10}$ charge.
A more direct way to get these IIA
solutions is simply to take the harmonic functions for the $D$=11
$M2$- or the $M5$-brane to be independent of one of the transverse
directions. A brane solution whose harmonic function
is independent of a number of transverse coordinates is
sometimes said to be ``delocalised", 
``averaged" or ``smeared" over those directions. Delocalised branes
will appear
when we discuss intersecting brane solutions.

\subsection{Intersecting $M$-branes}

We now turn to solutions corresponding to intersecting $M$-branes.
We begin by presenting the generalized supersymmetric solution 
for two $M2$-branes orthogonally ``overlapping" in a 
point \cite{tseytlinharm,gkt} which we will denote
by $M2\perp M2(0)$:
\bea
d{s}^2&=&\left(H_1H_2\right)^{1/3}
[-\left(H_1H_2\right)^{-1}dt^2 +
H_1^{-1}\left(dx_1^2+dx_2^2\right) +
H_2^{-1}\left(dx_3^2+dx_4^2\right)\nonu\\
&&\qquad\qquad+
\left(dx_5^2+\dots +dx_{10}^2\right)],\nonu \\
{F}_{t12\alpha}& =& {c_1\over 2}{\partial_\alpha H_1\over H_1^2}, \qquad
{F}_{t34\alpha} = {c_2\over 2}
{\partial_\alpha H_2\over H_2^2}, \qquad \a=5,\dots, 10.\nonu \nonu\\
H_i&=& H_i(x_5,\dots ,x_{10}), \qquad \nabla^2H_i=0,\qquad c_i=\pm1,\qquad
i=1,2.
\label{eq:m2m2}
\eea
There are Killing spinors of the form
$\epsilon=(H_1 H_2)^{-1/6}\eta$, where $\eta$ is constant and satisfies
the algebraic constraints
\bea
\hat\Gamma_{012}\eta&=&c_1\eta\nonu\\
\hat\Gamma_{034}\eta&=&c_2\eta.
\eea
Since $[\hat\Gamma_{012},\hat\Gamma_{034}]=0$
and ${\rm Tr}(\hat\Gamma_{012})(\hat\Gamma_{034})=0$,
each condition projects out an independent half of the
spinors and 
we conclude that there are eight Killing spinors and hence the solution
preserves 1/4 of the supersymmetry.

The functions $H_i$ are harmonic in the coordinates
$\vec x=\{x_5,\dots,x_{10}\}$ and we first take them to be of the form
\beq
H_i = 1 + {a_{i}\over r_{i}^4},\qquad
r_{i}= \left |\vec{x}-\vec{x}_{i}\right |. 
\label{eq:harm}
\eeq
The solution then describes an $M2$-brane oriented in the $\{1,2\}$ plane with
position $\vec{x}_{1}$ and another oriented in the $\{3,4\}$ plane with
position $\vec{x}_2$ orthogonally overlapping in a point.
To see this we note that the solution is a kind of superposition
of each individual $M2$-brane solution. For the directions
tangent to the $ith$ $M2$-brane the metric appears with
the inverse of the harmonic function i.e., $H_i^{-1}$, and the
directions transverse to the $ith$ $M2$-brane are independent of
$H_i$ exactly as in (\ref{eq:three}). Moreover,
the overall conformal factor is the product of the
two harmonic functions to the appropriate power, as one expects
for $M2$-branes.
In the degenerate case that
$\vec{x}_{1}=\vec{x}_{2}$, an $M2$-brane 
with $\{1,2\}$ orientation intersects one with $\{3,4\}$ orientation.
Note that the special case when $H_1=H_2$ 
was first constructed by G\"uven \cite{guven}\footnote{This
was described as a ``4-brane" solution in \cite{guven}
because of the $SO(4)$ invariance
in the $(x_1,x_2,x_3,x_4)$ plane. The problem with this interpretation is 
the absence of boost invariance
that single branes possess and it is best interpreted as a special
case of the $M2\perp M2(0)$ solution.}.

A more general solution has harmonic functions of the form
\beq
H_i = 1 + \sum_{I=1}^{k_i}{a_{i,I}\over r_{i,I}^4},\qquad
r_{i,I}= \left |\vec{x}-\vec{x}_{i,I}\right |. 
\eeq
The solution then
describes $k_1$ parallel $M2$-branes with $\{1,2\}$ orientation and
positions $\vec{x}_{1,I}$, and $k_2$ parallel $M2$-branes with $\{3,4\}$
orientation and
positions $\vec{x}_{2,I}$. Each $M2$-brane of one set orthogonally overlaps
all of the $M2$-branes in the other set in a point.
An $M2$-brane with $\{1,2\}$ orientation intersects 
one with $\{3,4\}$ orientation in
the case that
$\vec{x}_{1,I}=\vec{x}_{2,J}$, for some combination $I,J$.
Note that in describing the solutions in the rest
of the paper we will implicitly
take the harmonic functions to be that of a single brane as in
(\ref{eq:harm}) for ease of exposition. 

There is a potentially confusing point 
with our interpretation of (\ref{eq:m2m2}).
To explain
this lets first introduce some terminology:
we refer to
{\it common tangent} directions as being tangent directions common
to all
branes. In the case that the branes intersect rather than overlap
these are the intersection directions.
{\it Relative transverse} directions are those tangent to
at least one but not
all
branes and {\it overall transverse} directions are those orthogonal to all
branes. 
The two harmonic functions in (\ref{eq:m2m2})
are invariant under the common tangent direction, i.e., the time direction
in this case, and also under translations in all the relative transverse 
directions $x_1,\dots, x_4$. In particular, we note that
$H_1$ does not fall off in the $x_3,x_4$ directions, as one would
expect for
a $D$=11 $M2$-brane spatially oriented in the $\{1,2\}$ plane. i.e., the $H_1$
$M2$-brane is delocalised in the directions tangent to the other $M2$-brane.
Similarly the $H_2$ $M2$-brane in the $\{3,4\}$ plane is delocalised
in the directions tangent to the $M2$-brane lying in the $\{1,2\}$ plane. 
It is natural to conclude that our interpretation of the solutions 
as describing intersecting branes 
is valid but that we have not found the most general fully localised solutions.
In a later subsection
we will discuss more
general solutions that make some progress in this direction.

Since the $M2$-branes are delocalised in the directions tangent to
the other brane, we can immediately consider the solution 
(\ref{eq:m2m2}) in a dimensionally reduced context
with all relative transverse directions
periodically identified. This implies, e.g., that
the $M2$-brane with spatial orientation in the $\{1,2\}$ plane 
has been reduced in the $\{3,4\}$ directions to give a membrane in
$D$=9 and then wrapped
in the $\{1,2\}$ directions
to give a point object in $D$=7 that carries electric
charge of the $D$=7 gauge field $A_{\mu 12}$.
Similarly the other $M2$-brane is a point object in $D$=7
carrying electric charge with respect to the gauge field
$A_{\mu 34}$. Thus, the dimensionally reduced solution
may be regarded as two charged $D$=7 black
holes,
each carrying an electric charge with respect to different $U(1)$'s.
In the
intersecting case with $\vec x_1=\vec x_2$
the two black holes are coincident
and we can interpret it as a single black hole that carries two
charges. These BPS 
black holes solutions are extremal and in fact have
naked singularities. Later we will describe how extremal black
holes with non-zero horizon area can be constructed from intersecting
branes.

The solutions (\ref{eq:m2m2}) are generically singular on the surfaces
$\vec{x}-\vec{x}_{i}=0$, with the scalar curvature diverging.
This behavior is different from that of a single $M2$-brane where, as
we have noted, these
surfaces are
regular event horizons.
The singularity in the present case arises because the
$M2$-branes are delocalised
in the relative transverse dimensions.
It is possible that 
more general localised solutions will exhibit a similar 
singularity structure to that of a single $M2$-brane.

Lets now turn to configurations involving $M5$-branes. We will
present a solution describing an $M2$-brane intersecting an 
$M5$-brane in a one-brane, $M2\perp M5(1)$, and another 
describing an $M5$-brane intersecting another $M5$-brane 
in a threebrane,  $M5\perp M5(3)$. 
Both solutions are constructed as a kind of
superposition of their constituents.
There is another solution
involving $M5\perp M5(1)$ which is qualitatively different and
will be discussed in a later subsection.
The $M2\perp M5(1)$ solution is 
given by \cite{tseytlinharm,gkt}
\bea
d{s}^2 &=&H_1^{2/3}H_2^{1/3}[H_1^{-1} H_2^{-1}(-dt^2 +dx_1^2)+
H_1^{-1}\left(dx_2^2+dx_3^2+dx_4^2+dx_{5}^2\right)\nonu\\ && 
\qquad\qquad+
H_2^{-1}\left(dx_6^2\right)+
\left(dx_7^2+dx_8^2+dx_9^2+dx_{10}^2\right)],\nonu\\
{F}_{6\a\b\g}&=&{c_1\over 2}\epsilon_{\a\b\g\de}\partial_\de H_1,\qquad
{F}_{t16\a}={c_2\over 2}{\partial_\a H_2\over H_2^2},\qquad
H_i=H_i(x_7,\dots,x_{10}),
\label{eq:m2m5}
\eea
where $\epsilon_{\a\b\g\de}$ is the $D$=4 flat space alternating symbol.
The eight Killing spinors have the form $\epsilon=H_1^{-1/12}H_2^{-1/6}\eta$
with the constant spinor $\eta$ satisfying
\bea
\hat\Gamma_{016}\eta&=&c_1\eta\nonu\\
\hat\Gamma_{012345}\eta&=&c_2\eta.
\eea
and we have used the fact that $\hat\Gamma_{10}=\hat\Gamma_0\hat\Gamma_1\dots
\hat\Gamma_9$. If we choose the harmonic functions to have single 
coincident centres
then the solution describes an 
$M5$-brane 
in the $\{1,2,3,4,5\}$ direction intersecting an $M2$-brane in the $\{1,6\}$
direction.

The solution corresponding to $M5\perp M5(3)$ is given by
\cite{paptown,tseytlinharm,gkt}
\bea
d{s}^2 &=&\left(H_1H_2\right)^{2/3}[
\left(H_1H_2\right)^{-1}(-dt^2 +dx_1^2 +dx_2^2
+dx_{3}^2)+
H_1^{-1}\left(dx_4^2+dx_5^2\right)\nonu\\ && \qquad\qquad+
H_2^{-1}\left(dx_6^2+dx_7^2\right)+
\left(dx_8^2+dx_9^2+ dx_{10}^2\right)], \nonu\\
{F}_{67\a\b}&=&{c_1\over 2}\epsilon_{\a\b\g}\partial_\g H_1,\qquad
{F}_{45\a\b}= {c_2\over 2}\epsilon_{\a\b\g}\partial_\g H_2,
%\qquad\a=7,8,9,
\qquad H_i=H_i(x_8,x_9,x_{10}),
\label{eq:m5m5}
\eea
where $\epsilon_{\a\b\g}$ is the $D$=3 flat space alternating symbol.
The solution preserves $1/4$ of the supersymmetry and the Killing
spinors are given by $\epsilon=(H_1 H_2)^{-1/12}\eta$ with the constant
spinor $\eta$ satisfying the constraints
\bea
\hat\Gamma_{012345}\eta&=&c_1\eta\nonu\\
\hat\Gamma_{012367}\eta&=&c_2\eta.
\eea
If we choose the harmonic functions to have single coincident centres
then the solution describes an $M5$-brane in the $\{1,2,3,4,5\}$ direction
intersecting an $M5$-brane in the $\{3,4,5,6,7\}$ direction.

Note that in both solutions (\ref{eq:m2m5}), (\ref{eq:m5m5}) 
the harmonic functions again just depend on the
overall transverse directions.
Thus,  just
as in the $M2\perp M2(0)$ solution above, each of the branes are
delocalised along the directions tangent to the other. 
We will see later how the solutions (\ref{eq:m2m5}) and (\ref{eq:m5m5}) can
be obtained from (\ref{eq:m2m2}) after dimensional reduction, duality
transformations and then uplifting back to $D$=11.

\subsection{Multi-Intersections and Black Holes}

In the last section we presented three basic intersections of
two $M$-branes, each preserving 1/4 of the supersymmetry. We
can construct solutions of $n$ orthogonally intersecting
$M$-branes by simply ensuring that the branes are aligned along
hyperplanes in such a way that the pairwise intersections 
are amongst the allowed set. The solutions are then constructed by
superposing the solutions in a way that we have already seen:  there is
a harmonic function $H$ for each constituent brane that depends
on the overall transverse coordinates. It appears in the
metric only as $H^{-1}$ multiplying the directions tangent to that
brane and in the overall conformal factor with the appropriate power
depending on whether it is an $M2$- or an $M5$-brane. The four-form
field strength has non-zero components corresponding to those of
each of the $M$-branes. This procedure
\cite{tseytlinharm,gkt} has been called the ``harmonic function rule".

Generically a configuration of $n$ intersecting branes
will preserve $2^{-n}$ of the supersymmetry \cite{paptown,tseytlinharm,gkt}. 
This is because
the Killing spinors are projected out by products of
Gamma matrices 
with indices tangent to each brane, and generically these projections
are independent.
There are, however,  
important exceptions when the projections are not independent
\cite{klebtseyt,gkt}. 
Let us illustrate this
by discussing the cases for three intersecting $M$-branes which all
preserve $1/8$ of the supersymmetry.
There is a unique configuration corresponding
to three  
$M2$-branes. If the $M2$-branes are orientated 
along the $\{1,2\}$, $\{3,4\}$ and $\{5,6\}$ 
hyperplanes the metric is given by \cite{tseytlinharm,gkt}:
\bea
ds^2&=&(H_1H_2H_3)^{1/3}[-{\left(H_1H_2H_3\right)}^{-1}dt^2
+H_1^{-1}\left(dx_1^2 + dx_2^2\right)
+H_3^{-1}\left(dx_3^2 + dx_4^2\right)\nonu\\
&&\qquad\qquad+H_3^{-1}\left(dx_5^2 + dx_6^2\right)
+\left(dx_7^2+dx_8^2+dx_9^2+dx_{10}^2\right)],
\label{eq:twotwotwo}
\eea
with the harmonic functions $H_i=H_i(x_7,x_8,x_9,x_{10})$.

There is also a unique configuration corresponding to two
$M2$-branes and one $M5$-brane and we can take the
%which preserves $1/8$ of the
%supersymmetry. 
orientations of the branes to be in the $\{1,2,3,4,5\}$,
$\{1,6\}$ and $\{2,7\}$ hyperplanes. 
This solution provides us with the first special
triple intersection. To see this note that the product of the three
Gamma matrix projections gives another projection 
corresponding to an $M5$-brane in the $\{3,4,5,6,7\}$ direction. This means 
that we can obtain an $M2\perp M2\perp M5\perp
M5$ configuration that breaks $1/8$ of the 
supersymmetry (and not $1/16$ 
as one might naively expect) as long as we choose the 
polarisation of the fourth $M5$-brane 
(i.e., whether it is a brane or anti-brane) to be determined by
the polarisations of the first three.
The metric for this solution is given by \cite{klebtseyt}
\bea
&&ds^2=(H_1H_2)^{1/3}(H_3H_4)^{2/3}[-{\left(H_1H_2H_3H_4\right)}^{-1}dt^2
+(H_1H_3)^{-1}\left(dx_1^2\right)
+(H_2H_3)^{-1}\left(dx_2^2\right)\nonu\\
&&\qquad\qquad+(H_3H_4)^{-1}\left(dx_3^2+dx_4^2+dx_5^2\right)
+(H_1H_4)^{-1}\left(dx_6^2\right)
+(H_2H_4)^{-1}\left(dx_7^2\right)\nonu\\
&&\qquad\qquad+\left(dx_8^2+dx_9^2+dx_{10}^2\right)],
\label{eq:twotwofivefive}
\eea
with the harmonic functions $H_i=H_i(x_8, x_9, x_{10})$.

There are two ways in which two $M5$-branes and one $M2$-brane can 
intersect. The first is when they are oriented along the
$\{1,2,3,4,5\}$, $\{3,4,5,6,7\}$ and $\{1,6\}$ planes. Note that this
is again a special intersection as we can add an $M2$-brane
in the $\{2,7\}$ plane to return to the solution (\ref{eq:twotwofivefive}).
The other intersection has the $M2$-brane lying in the $\{3,8\}$
plane and the three branes intersect in a common string. 
For this solution there are only two
overall transverse directions and so the three harmonic functions have 
logarithmic divergences. 

Finally there are three ways in which three $M5$-branes can intersect.
Take the first two to lie in the $\{1,2,3,4,5\}$ and 
$\{3,4,5,6,7\}$ planes. The third $M5$-brane can be placed 
in the $\{1,2,3,6,7\}$ direction in which case 
there is an overall string intersection. We shall 
return to this configuration in a moment. 
If the third $M5$-brane is placed in the $\{1,3,4,6,8\}$
plane there is a common 2-brane intersection and
we obtain a third special triple intersection since we can add 
a fourth $M5$-brane in the $\{2,3,4,7,8\}$ plane and still preserve
$1/8$ of the 
supersymmetry. Note that this configuration has only two overall
transverse dimensions. The third case has the $M5$-brane lying in the
$\{3,4,5,8,9\}$ plane and now there is only one overall transverse dimension. 

Although conceptually clear it is slightly involved to
list all of the supersymmetric intersecting $M$-brane configurations
and determine the amount of supersymmetry preserved taking into
account the three special triple intersections. This was undertaken in
\cite{bergmult}.

We now turn to intersecting brane configurations corresponding to
BPS black holes in $D$=4,5 that have non-zero horizon
area. To obtain such a black hole in $D$=5 we can dimensionally reduce the
$M2\perp M2\perp M2$ solution (\ref{eq:twotwotwo}) 
along the six relative transverse
directions $x_1,\dots,x_6$.
If we take the harmonic functions $H_i$ to
have a single coincident centre we 
are led \cite{tseytlinharm} to a 
black hole solution in $D$=5 that carries three electric
charges corresponding to three $U(1)$'s coming from the three-form
components $A_{\mu12}$,  $A_{\mu34}$ and $A_{\mu56}$.
One can show that the BPS black hole is extremal and has non-zero horizon area.
There is another way to obtain such a $D$=5 black hole. One considers
the $M2\perp M5(1)$ solution (\ref{eq:m2m5})
and adds momentum along the string
direction. The procedure for doing this is well known and the
solution one gets is \cite{tseytlinharm}
\bea
d{s}^2 = &&H_1^{2/3}H_2^{1/3}[H_1^{-1} H_2^{-1}(dudv +Kdu^2)+
H_1^{-1}\left(dx_2^2+dx_3^2+dx_4^2+dx_{5}^2\right)\nonu\\ 
&&\qquad\qquad +
H_2^{-1}\left(dx_6^2\right)+
\left(dx_7^2+dx_8^2+dx_9^2+dx_{10}^2\right)],
\label{eq:firstm2m5wave}
\eea
where $u,v=x_1\pm t$ and the function $K$ is harmonic in the overall
transverse coordinates: in the simplest case of a single centre it
corresponds to a ``pp-wave" carrying momentum in the string direction.
The wave 
in the $\{1\}$ direction imposes the constraint
\beq
\epsilon=\pm\hat\Gamma_{01}\epsilon
\eeq
on the Killing spinors ($\pm$ depending on which direction it is travelling).
It thus breaks a further 1/2 of the supersymmetry and hence the
solution preserves 1/8 of the supersymmetry. Reducing this to $D$=5 along
the relative transverse directions and the string intersection,
we obtain a black hole that carries electric $A_{\mu16}$ charge,
magnetic $A_{\mu\nu6}$ charge (note that in $D$=5 this is dual to
a vector field) and electric Kaluza-Klein $g_{\mu1}$ charge corresponding to
the momentum running along the string. 

Let us now discuss how $D$=4 black holes can be constructed from 
intersecting $M$-branes. One way is to dimensionally reduce the
$M2\perp M2\perp M5\perp M5$ solution (\ref{eq:twotwofivefive})
along the relative transverse
directions \cite{klebtseyt}. In this way one obtains a black hole carrying
two electric and two magnetic charges.
Another way is to consider three
$M5$-branes all overlapping in a common string, with momentum
running along the common string direction \cite{klebtseyt}:
\bea
ds^2&=& (H_1H_2H_3)^{2/3}\big[(H_1H_2H_3)^{-1}\left(dudv+Kdu^2\right)
+(H_1H_2)^{-1}\left(dx_2^2+dx_3^2\right)\nonu\\
&+&(H_1H_3)^{-1}\left(dx_4^2+dx_5^2\right)
+(H_2H_3)^{-1}\left(dx_6^2+dx_7^2\right)
+\left(dx_8^2+\dots dx_{10}^2\right)].
\label{eq:fivefivefivewave}
\eea
Note that as long as the direction of the wave is chosen appropriately,
it does not 
impose any additional constraints on the Killing spinors and hence the
solution preserves 1/8 of the 
supersymmetry.

It should become clear in the next section that
the different configurations of $M$-branes giving black holes in 
either $D$=4 or $D$=5
can be related to each other by dimensional reduction and duality.
There we will also discuss ways in which intersecting $D$-branes give rise to
black holes. The perturbative $D$-brane point view
has been very successfully exploited in giving a microscopic
interpretation of black hole entropy. As less is understood about $M$-brane
dynamics it is harder to do this in $M$-theory. 
However, one can turn this around and see what can be learned about $M$-theory
dynamics if we demand that it is consistent with black hole entropy.
This has been pursued in \cite{klebtseyt}.

\subsection{Dynamics of Intersections}

As we have noted all of the solutions we have considered
so far are delocalised along the relative transverse
directions i.e., in the directions tangent to all of the branes.
As such, the properties and dynamics of the intersection are 
somewhat occluded.  
Addressing this directly at the level of finding 
more general solutions is an interesting open question 
but we can also obtain a great deal of insight using more
general arguments \cite{stromopen,townsurg}.

Lets begin by considering the possibility 
of an $M2$-brane ending on an $M5$-brane in
a string. One immediately faces a potential problem with
charge conservation: consider a seven-sphere surrounding
the $M2$-brane. The integral of $*F$, 
along this seven sphere gives the $M2$-brane charge $Q_e$,
where $F$ is the four-form
field strength.
It might seem that we could 
smoothly deform the sphere to a point by slipping 
it off the end past the $M5$-brane and hence conclude that
$Q_e$ must vanish. However, this argument ignores
what happens when the sphere is passed through the $M5$-brane. 
The argument can fail if the charge can somehow be carried 
by the string boundary inside the $M5$-brane.

One way to study this is to include the world-volume dynamics 
of the $M5$-brane in the supergravity equations of motion.
The low-energy dynamics of an $M5$-brane and its coupling
to the spacetime supergravity fields can be described by a low-energy
effective action on the world-volume of the brane. This can be constructed
from first principles by determining the zero modes in the small fluctuations
around a classical solution. 
The dynamics for the $M5$-brane is
governed by a $D$=6 $(0,2)$ supermultiplet 
multiplet whose bosonic fields consist of
5 scalars and a two-form $V_2$ that has self dual field
strength \cite{chsbrane}. The world-volume action contains the coupling
$|dV_2 - A|^2$ where $A$ is the supergravity three-form pulled back 
to the world-volume: $A_{ijk}$=$A_{\mu\nu\rho}\p_iX ^\mu\p_j X^\nu \p_k X^\rho$,
where $X^\mu(\sigma^i)$ are the world-volume scalar coordinates. 
This modifies the $A$ equation of motion to
include a world-volume source term.  After integrating over an asymptotic
seven sphere we 
deduce that $Q_e=\int_{S^3} *dV_2$ where the integral is a world-volume
integral and $*$ is the world-volume
Hodge-dual. In the world-volume theory this integral is non-zero if
there is a self-dual string inside the six-dimensional world-volume. 
Thus we conclude that
it is possible for an $M2$-brane to end in a string on an $M5$-brane if
the $M2$-brane charge is carried by a self-dual string inside the world-volume
theory. 
Note that it is also possible to reach 
an identical conclusion without having to introduce world-volume
dynamics if one takes into account the contribution of Chern-Simons
couplings in the supergravity \cite{townsurg}.

This conclusion indicates that the $M5$-brane is a natural generalisation 
of a $D$-brane
in string theory to $M$-theory. 
%Instead of a fundamental string ending
%on a point on the world-volume of the $D$-brane.
It also suggests that
we can think of the $M2\perp M5(1)$ solution (\ref{eq:m2m5}) as being 
associated with
these configurations. It is possible that more general supergravity
solutions exist that have localised $M2$-branes ending on $M5$-branes.
They would be very interesting as
they would illuminate the geometry of the boundary of the $M2$-brane and
the dynamics of the self-dual string. 
These solutions will 
probably
be highly non-trivial to construct
but perhaps progress can be made
by looking for localised solutions with an $M2$-brane ending on
the $M5$-brane from either side.

Similar arguments can be developed for self intersections 
of $M$-branes.
The following argument in fact
works for all $p$-branes \cite{paptown}. 
If we assume that we can consider a $q$-brane
intersection within a given $p$-brane as a dynamical object in the
$p+1$-dimensional world-volume field theory, then  
the condition that the $p$-brane
can support a dynamical $q$-brane intersection would be that its world
volume contains a $(q+1)$-form potential to which the $q$-intersection
can couple. The effective action of all $p$-branes contain scalar fields
which are the Goldstone modes arising from the fact that the classical
$p$-brane solution breaks translation invariance. These scalar fields
have one-form field strengths which can be dualised in the world-volume
to give 
$(p-1)$-form dual potentials which can couple to
a $q=(p-2)$-dimensional intersection. Hence we conclude that a $p$-brane
can have a dynamical self intersection in $(p-2)$ dimensions. 
The $M2\perp M2(0)$ and $M5\perp M5(3)$ solutions (\ref{eq:m2m2}),
(\ref{eq:m5m5}) are both 
consistent with this rule.

\subsection{$M5\perp M5(1)$}

There is another solution corresponding to two $M5$ branes
overlapping in a string \cite{gkt}:
\bea
d s^2&=&\left(H_1H_2\right)^{2/3}[
(H_1H_2)^{-1}(-dt^2+dx_1^2)+
H_2^{-1}(dx_2^2+dx_3^2+dx_4^2+dx_5^2)\nonu\\
&&\qquad\qquad+H_1^{-1}(dx_6^2+dx_7^2+dx_8^2+dx_9^2)
+dx_{10}^2]\nonu\\
F_{mnp10}&=&-{c_1\over 2}\epsilon_{mnpq}\p_qH_1,\qquad
 F_{\mu\nu\lambda10}= -{c_2\over 2}\epsilon_{\mu\nu\lambda\rho}
\p_\rho H_2,\nonu\\
H_1&=&H_1(X^1_m),\qquad H_2=H_2(X^2_\mu),\qquad \nabla^2H_i=0,
\label{eq:m5m5over}
\eea
where $X^1_m=(x_2,x_3,x_4,x_5)$ and 
$X^2_\mu=(x_6,x_7,x_8,x_9)$. 
For single centre harmonic functions this 
corresponds to an $M5$-brane with orientation 
$\{1,2,3,4,5\}$ 
overlapping another with orientation $\{1,6,7,8,9\}$. There are
16 Killing spinors of the form $\epsilon=(H_1H_2)^{-1/12}\eta$
with the constant spinor $\eta$ satisfying
\bea
\hat\Gamma_{016789}\eta&=&c_1\eta\nonu\\
\hat\Gamma_{012345}\eta&=&c_2\eta,
\label{eq:proj}
\eea
It satisfies the harmonic function rule but with a key difference:
the harmonic functions are now independent of the single overall
transverse direction and only depend on the 
relative transverse directions. 
That is, the $M5$-branes are now localised inside
the directions tangent to the other $M5$-brane but are delocalised in
the overall transverse direction that separates them.

Another interesting feature of this solution is that it does
not satisfy the $(p-2)$ dimensional self-intersection rule for
$p$-branes that we discussed in the last subsection.
The resolution of this puzzle is quite interesting.
A consequence of the Gamma-matrix projections (\ref{eq:proj}) 
is that $\hat\Gamma_{0110}\eta=c_1c_2\eta$.
This suggests that we can add an $M2$-brane in the $\{1,10\}$
plane without breaking further supersymmetry.
Note that such an $M2$-brane overlaps each of the 
$M5$-branes in a string which is allowed. The solution is
given by \cite{tseytlincomptwo,jpgunpub}
\bea
d s^2&=&\left(H_1H_2\right)^{2/3}H_3^{1/3}[
(H_1H_2H_3)^{-1}(-dt^2+dx_1^2)+
H_2^{-1}(dx_2^2+dx_3^2+dx_4^2+dx_5^2)\nonu\\
&&\qquad\qquad+H_1^{-1}(dx_6^2+dx_7^2+dx_8^2+dx_9^2)
+H_3^{-1}dx_{10}^2]\nonu\\
F_{mnp10}&=&-{c_1\over 2}\epsilon_{mnpq}\p_qH_1,\qquad
 F_{\mu\nu\lambda10}= -{c_2\over 2}\epsilon_{\mu\nu\lambda\rho}
\p_\rho H_2
\qquad F_{t110I}={c_1c_2\over 2}{\p_I H_3\over H_3^2},
\label{eq:only}
\eea
where $x_I=(X^1_m,X^2_\mu)$ and the function $H_3(X^1,X^2)$ 
corresponding to the 
$M2$-brane satisfies the equation
\beq
\left[H_1^{-1}(X^1)\nabla^2_{(X^1)}+H_2^{-1}(X^2)\nabla^2_{(X^2)}\right]H_3=0.
\label{eq:onenext}
\eeq
Functions of the form
\beq
H_3(X^1,X^2)= h_1(X^1) + h_2(X^2),
\label{eq:onep}
\eeq
solve this equation if the $h_i$ are harmonic on $\bE^4$, but point
singularities of $h_1$ or $h_2$ would represent $M2$-branes that are delocalized
in four more directions. We expect that there exist solutions of 
(\ref{eq:onenext})
representing localized $M2$-branes 
although explicit solutions may be difficult to find. 
In the same way the solution $M2\perp M5(1)$ can be thought of
as being related to an $M2$-brane ending on an $M5$-brane we can think
of 
the solution (\ref{eq:only}) as corresponding
to an $M2$-brane being stretched between two $M5$-branes. This
interpretation and the fact that we can add the extra $M2$-brane
without breaking any more supersymmetry also provides a
resolution of the fact that the solution (\ref{eq:m5m5over}) violates
the $(p-2)$ self-intersection rule: when two $M5$-branes are brought
together to intersect on a string, one should think of the intersection
as being a collapsed $M2$-brane.

This observation suggests the following nomenclature: the
solutions $M2\perp M2(0)$, $M2\perp M5(1)$ and $M5\perp M5(3)$
can be called intersecting brane solutions, since when they do intersect
(as opposed to overlap) they describe dynamical intersections. On
the otherhand the $M5\perp M5(1)$ solution should be described as
an overlap since it is not until we add an extra $M2$-brane that
we get a dynamical intersection. 

It is worth noting that if we remove one of the $M5$-branes in 
(\ref{eq:only}) we obtain a more general solution than the
previous $M2\perp M5(1)$ solution (\ref{eq:m2m5})
in that the equation for the
$M2$-brane coming from (\ref{eq:onenext}) is more general than 
just a harmonic function in the overall transverse coordinates.
Again we do not know of any interesting solutions in closed form.
There are also generalisations of the $M2\perp M2(0)$ and $M5\perp M5(3)$
solutions where one of the $M$-branes satisfies a more general 
equation. These can be obtained by dimensional reduction 
and duality using the results of the next section.
Finally we note that more general configurations of multi-intersecting
$M$-branes can be obtained by combing these types of intersections with
the previous ones. See \cite{bergmult} for some results in this direction.

%%%%%%%%%%%%%%%%%%%%%%%
\section{Intersecting Branes in Type II String Theory}
%%%%%%%%%%%%%%%%%%%%%%%%

%The dimensional reduction of $D$=11 supergravity gives
%the type IIA supergravity which is the low-energy
%effective action of
%type IIA string theory. 
%By dimensionally reducing the $M$-brane solutions considered in
%the previous section we obtain intersecting brane solutions of the
%type IIA theory. We can then use $T$-duality and $S$-duality to generate
%additional solutions of both the type IIA and IIB theory. In this
%section we outline how this is achieved.

\subsection{$NS$ and $D$-branes}

The $D$=10 type IIA supergravity action can be obtained from 
dimensional reduction on a circle of $D$=11 supergravity.
The Kaluza-Klein ansatz for the
bosonic fields leading to the string-frame 10-metric is
\bea
ds^2_{(11)} &=& e^{-{2\over3}\phi(x)}dx^\mu dx^\nu g_{\mu\nu}(x) +
e^{{4\over3}\phi(x)}\big( dy+ dx^\mu C_\mu(x)\big)^2 \nonu\\
A_{(11)} &=& A(x) + B(x)\wedge dy\ ,
\label{eq:twoe}
\eea
where $A_{(11)}$ is the $D$=11 three-form potential and $x^\mu$ are the $D$=10
spacetime coordinates. 
We read off from the right hand side the bosonic fields of
$D$=10 IIA supergravity; these are the
$NSNS$ fields $(\phi, g_{\mu\nu}, B_{\mu\nu})$ and the  $RR$
fields $(C_\mu, A_{\mu\nu\rho})$. 
The bosonic fields of the $D$=10 type IIB supergravity 
coming from the $NSNS$ sector are identical to that of
the type IIA theory, $(\phi, g_{\mu\nu}, B^{(1)}_{\mu\nu})$.
From the $RR$ sector of the IIB theory 
there is an axion, another two-form and
a four-form that has a self-dual field strength $(l, B^{(1)}_{\mu\nu},
A^{+}_{\mu_1\mu_2\mu_3\mu_4})$.

The rank of the various form potentials 
immediately suggests what the spectrum
of BPS branes is. A potential of rank 
$r$ has a field strength of rank $(r+1)$ that
can be integrated along an $(r+1)$-sphere
which in $D$ spacetime dimensions surrounds a $(D-3-r)$-brane.
The value of the integral gives the magnetic $r$-form charge
carried by the the $(D-3-r)$-brane.
Similarly, the field strength
of rank $(D-1-r)$ that is Poincare dual to the $(r+1)$-form 
field strength can be integrated along
a $(D-1-r)$ sphere that surrounds an $(r-1)$-brane. 
Now the integral gives the electric $r$-form charge carried
by the $(r-1)$-brane.
Of course one still needs to check that such solutions to the non-linear
field equations exist and moreover to check if they admit any Killing spinors.
This has been carried out and we record here 
the metric and dilaton behaviour of the various BPS solutions.

The IIA and IIB $NS$-strings are 
electrically charged with respect to the $NS$
two-form. For both the type IIA and IIB theory we have:
\bea
ds^2&=&H^{-1}\left(-dt^2+dx_1^2\right)+dx_2^2+\dots+dx_9^2\nonu\\
e^{2\phi}&=&H^{-1},\qquad H=H(x_2,\dots,x_9),\qquad \nabla^2H=0. 
\label{eq:fundstringsol}
\eea
The IIA and IIB $NS5$-branes carry magnetic $NS$ two-form charge and
we have:
\bea
ds^2&=&-dt^2+dx_1^2+\dots+ dx_5^2 +H\left(dx_6^2+\dots+dx_9^2\right)\nonu\\
e^{2\phi}&=&H,\qquad H=H(x_6,\dots,x_9),\qquad \nabla^2H=0. 
\label{eq:nsfivesol}
\eea
Finally, 
the $Dp$-branes carry either electric or magnetic charge
with respect to the $RR$ fields and the metric and dilaton are given by:
\bea
ds^2&=&H^{-1/2}\left(-dt^2+dx_1^2+
\dots +dx_p^2\right)+H^{1/2}\left(dx_{p+1}^2+\dots
+dx_9^2\right),\nonu\\
e^{2\phi}&=&H^{-{(p-3)\over 2}},\qquad H=H(x_{p+1},\dots,x_9),
\qquad \nabla^2H=0. 
\label{eq:dbranesol}
\eea
Given the rank of the $RR$-forms that we mentioned above,
we see that the type IIA
theory has $Dp$-branes with $p=0,2,4,6$. There is an additional
$D8$-brane which is related to massive type IIA supergravity and
we refer the reader to \cite{bdgpt} for more details. For the IIB
theory we have $p=-1,1,3,5,7$. $p=-1$ corresponds to an instanton 
\cite{GGP} and
we won't include it in our discussions of intersecting branes. 
Note that we have written all of the above solutions in the 
sigma-model string metric
which is related to the Einstein metric via $g_E=e^{-\phi/2}g_\sigma$.

All of these type II branes preserve 1/2 of the supersymmetry.
The type II theories have two spacetime supersymmetries
parameters given by Majorana-Weyl spinors $\ep_L,\ep_R$.
In the type IIA theory they have opposite chirality and we
choose $\Gamma_{10}\ep_L=\ep_L$ and $\Gamma_{10}\ep_R=-\ep_R$. 
In the type IIB theory they have the same chirality and we choose
$\Gamma_{10}\ep_L,\ep_R=\ep_L,\ep_R$. 
The solutions have 16 Killing spinors which
satisfy the following projections:
\bea
{\rm IIA/IIB}\quad NS{\rm -strings}:\qquad&&\ep_L=\hat\Gamma_{01}\ep_L
\quad\qquad\ {}\ep_R=-\hat\Gamma_{01}\ep_R\nonu\\
{\rm IIA}\quad NS5{\rm-branes}:\qquad&&\ep_L=\hat\Gamma_{012345}\ep_L
\qquad\ep_R=\hat\Gamma_{012345}\ep_R\nonu\\
{\rm IIB}\quad NS5{\rm-branes}:\qquad&&\ep_L=\hat\Gamma_{012345}\ep_L
\qquad\ep_R=-\hat\Gamma_{012345}\ep_R\nonu\\
{\rm IIA/IIB}\quad 
Dp{\rm-branes}:\qquad&&\ep_L=\hat\Gamma_{01\dots p}\ep_R.
\label{eq:susyrules}
\eea
The Gamma-matrix projections will have an extra minus sign for the
corresponding anti-branes.

Let us first make a few brief comments on these different branes. 
The $NS$-string solutions that exist for each theory are simply identified
with the fundamental string of each theory \cite{dh,dghr,dghw,calmaldpeet}. 
The IIA and IIB 
$NS5$-branes of each theory are like solitons in quantum
field theory in the sense that their tension $T$ is related to the string
coupling $g$ and magnetic three-form
charge $Q$ via $T\sim Q/g^2$. These solitons have
an elegant $(4,4)$ superconformal field theory
description which is illuminating, but incomplete
\cite{chsworld}.
Although the IIA and IIB $NS5$-branes have the same supergravity 
solution the world-volume
theories that govern
the low-energy dynamics of these solitons are quite different \cite{chsbrane}.
The IIA $NS5$-brane has $(0,2)$ supersymmetry on the six-dimensional
world-volume just as the $M5$-brane. The bosonic fields consist of
five real scalars
and a two-form with self dual field strength.
The world-volume theory of the IIB $NS5$-brane has $(1,1)$ supersymmetry
whose bosonic field content is four scalars and a vector field.

The branes that carry charge with respect to the $RR$ fields are the 
$D$-branes. These branes differ from the $NS$-branes in that their tension 
is related to the string coupling and charge via $T\sim Q/g$. This fact
is closely related to the fact that $D$-branes have a very simple perturbative
description in string theory \cite{polch}. 
At weak coupling, they are surfaces in
flat spacetime where
open strings can end i.e., if we let $X^\mu$, $\mu=0,\dots,p$, 
be the coordinates tangent
and $X^T$, $T=p+1,\dots, 10$, be the coordinates transverse
to the brane,
then the strings coordinates $X^\mu(\tau,\sigma)$
satisfy Neumann boundary conditions and $X^T(\tau,\sigma)$ satisfy
Dirichlet boundary conditions. This perturbative description has played 
a central role in recent developments in string theory. 
Note that 
$D9$-branes fill all of space and a closer analysis leads
one to the type I theory. They are not associated with any supergravity
solution.
The world-volume theory for all $Dp$-branes is given by the dimensional
reduction of ten-dimensional superYang-Mills theory to $(p+1)$ dimensions.
The bosonic fields are $9-p$ 
scalars and a single vector field.

It will be convenient when we come to discussing intersecting brane
solutions to know how the above solutions are related. We noted
earlier that if we wrap the $M2$- or $M5$-brane on a circle we are led to
the type IIA $NS$-string and $D4$-brane, respectively, while if we reduce
the $M$-branes then we get the $D2$-brane and the $NS5$-brane,
respectively.
The second observation is that the type II brane solutions are
related by
various symmetries of the supergravity equations of motion.
The type IIB supergravity has an $SL(2,\bR)$ symmetry
of which an $SL(2,\bZ)$ is conjectured to survive as a non-perturbative 
symmetry of the string theory. The action on the low-energy fields
is as follows: the
$NSNS$ and $RR$ two-forms $B^{(i)}_{\mu\nu}$
transform as a doublet, the self dual four-form $A^+_{\mu_1\mu_2\mu_3\mu_4}$
and the Einstein metric are invariant
and the dilaton and $RR$ scalar can be packaged into a complex scalar 
$\tau=l+ie^{-\phi}$
which undergoes fractional liner transformations. 
The 
$Z_2$
``$S$-duality" transformation that interchanges the two-forms and acts as 
$\tau\to-1/\tau$ allows us to construct 
the $NS5$-brane and $NS$-string solutions
from the $D5$-brane and $D1$-brane solutions, respectively, and
vice-versa. 
Note that the $D3$-brane is left inert under this and all $SL(2,\bZ)$
transformations. For the behaviour of the $D7$-brane see 
\cite{GGP}.
If we employ more general $SL(2,\bZ)$ transformations then we
obtain ``non-marginal" BPS branes in the type IIB theory.  Specifically
if we start with a $NS5$-brane we obtain a
$(p,q)$ $5$-brane that is a bound state of $p$ $NS5$-branes and $q$
$D5$-branes, with $p$ and $q$ relatively prime integers. Similarly from the 
$NS$-string we get $(p,q)$ strings \cite{schwarzpq}. 
Since the $SL(2,\bZ)$ transformations do not break supersymmetry,
all of these solutions preserve $1/2$ 
of the supersymmetry: the projections on the Killing spinors are the
$SL(2,\bZ)$ rotations of those in (\ref{eq:susyrules}). 
The $(p,q)$ $5$-branes will play a role
when we discuss branes intersecting at angles in the next section.

The other basic tool to relate various branes is $T$-duality.
The type IIA theory compactified on a circle of radius $R$ is
$T$-dual to the IIB theory compactified on a circle of
radius $1/R$. This can be established exactly in perturbation theory.
Since $T$-duality interchanges Dirichlet with Neumann boundary conditions
\cite{polch}, 
if we perform $T$-duality in a direction transverse (tangent)
to a $Dp$-brane we obtain
a $D(p+1)$-brane ($D(p-1)$-brane). This can also be seen at the level
of classical supergravity solutions using the fact that
$T$-duality manifests itself as the ability 
to map a solution with an isometry into
another solution. The action of $T$-duality on the $NS$ fields with
respect to a
symmetry direction $z$, mapping
string-frame metric to string-frame metric,is  
\bea
d\tilde s^2 &=& \big[g_{\mu \nu}- g_{zz}^{-1}(
g_{\mu z}g_{z\nu}+B_{\mu z}B_{z\nu})\big] dx^\mu dx^\nu  
+ 2g_{zz}^{-1}B_{z\mu} dzdx^\mu +
g_{zz}^{-1} dz dz\nonu\\
\tilde B &=& {1\over2}dx^\mu\wedge dx^\nu \big[ B_{\mu \nu} -
g_{zz}^{-1}
(g_{\mu z}B_{z\nu}+B_{\mu z}g_{z\nu})\big] + 
g_{zz}^{-1} g_{z\mu} dz\wedge dx^\mu\nonu\\
\tilde\phi &=& \phi -{1\over2} \log g_{zz} 
\eea
where $x^\mu$ are the rest of the coordinates and
we have indicated the transformed fields by a tilde. 
These rules may be read as
a map either from IIA to IIB or vice-versa.
%symmetry direction $\alpha$, on
%the dilaton and string metric is simply given by 
%%
%\beq
%\bar{g}_{\alpha\alpha}=1/\hat{g}_{\alpha\alpha},\qquad
%e^{-2\bar{\varphi}}=
%\hat{g}_{\alpha\alpha}e^{-2\hat{\phi}}, 
%\eeq
%where $\hat \phi$ is the IIA dilaton and $\bar{\varphi}$ is
%the IIB dilaton.  
The action on the $RR$ gauge fields 
can be found in \cite{BHO}. 
%Let us note however,
%that off diagonal components of the metric $g_{\mu\alpha}$
%get interchanged with components $B_{\mu\alpha}$ of the $NS$
%two-form. 
For our applications the new solution obtained by $T$-duality
will preserve the same amount of supersymmetry as the original one.
Using these transformations we again conclude that if we perform $T$-duality
on a direction tangent to a $Dp$-brane solution (\ref{eq:dbranesol})
then we are led to a $D(p-1)$-brane solution. For example, the 
metric component $g_{zz}=H^{-1/2}\to H^{1/2}$.
But note that we do not arrive at the most general solution as the
harmonic function of the  $D(p-1)$-brane is invariant under the 
$z$ direction. Similarly, if we take a $Dp$-brane that
is delocalised in a transverse direction and perform $T$-duality
in that direction we get a $D(p+1)$-brane solution.
Performing $T$-duality on a direction transverse to 
a IIA/B  fundamental string 
(\ref{eq:fundstringsol}) delocalised in that direction 
will
transform it into a IIB/A fundamental string. Acting on a direction
tangent to the IIA/B string, say in the $\{1\}$ direction,
will replace the $B_{01}$ component of the $NS$ two-form
in the string solution with
an off diagonal term in the metric $g_{01}$. 
The final solution is a pp-wave of the IIB/A theory. 
Note that this is is the spacetime
manifestation of the fact that in perturbation theory $T$-duality
interchanges winding and momentum modes.
The supersymmetry projections for pp-waves travelling in a given direction
for both the IIA and IIB
theories are:
\beq
{\rm IIA/IIB}\quad {\rm pp-wave}:\qquad\ep_L=\hat\Gamma_{01}\ep_L
\quad\qquad\ {}\ep_R=\hat\Gamma_{01}\ep_R.
\eeq
Acting with $T$-duality in a direction tangent to a IIA/B $NS5$-brane 
(\ref{eq:nsfivesol}) will lead to
a IIB/A $NS5$-brane. If it is in a direction transverse to a
delocalised  
$NS5$-brane then we again 
get off diagonal terms in the metric. One finds that the non-trivial
part of the metric is
given by Taub-NUT space, which we will review in the next section.
These $T$-duality results are summarised in Table 1.
\begin{table}[h!]
\centering
\begin{tabular}{|>{$}c<{$}|>{$}c<{$}|>{$}c<{$}|}
\hline
& {\rm Tangent}& {\rm Transverse} \\
\hline
NS1&{\rm pp-wave} &NS1 \\
NS5& NS5& {\rm Taub-NUT}\\
Dp&D(p-1) &D(p+1)\\
\hline
\end{tabular}
\vspace{8pt}
\caption{T-Duality Rules For Type II Branes\label{tab:xxxx}}
\end{table}

Using these duality transformations we can essentially obtain all
others starting from, say the $M2$-brane. Reducing the $M2$-brane
to $D$=10 we obtain the IIA $D2$-brane. $T$-dualising this solution
leads to all $Dp$-brane solutions of the IIA and IIB theory. To
implement this for $p\ge 7$ one must use the massive IIA 
supergravity \cite{bdgpt}. On the otherhand
$S$-duality on the $D1$ and $D5$-branes gives the IIB $NS$-string and
the $NS5$-brane solutions, respectively. 
The corresponding IIA $NS$-branes are then obtained
by $T$-dualising on a transverse or tangent direction, respectively.
Similarly, we obtain the IIA/B pp-waves (Taub-NUT) from the IIB/A
$NS$-string ($NS5$-brane) by $T$-dualising on a tangent (transverse)
direction.
The $M5$-brane solution can be obtained by ``uplifting" either the 
$D4$-brane or the IIA $NS5$-brane to $D$=11. 
Uplifting the IIA $D0$-brane gives a $D=11$ pp-wave and, as we shall discuss in
the next section, uplifting the $D6$-brane gives Taub-NUT space.
Note that in performing these transformations we will be led to
the correct BPS solutions 
but possibly not the most general solution
as the harmonic function may become delocalised in the procedure, as we noted
above.

%%%%%%%%%%%%%%%%%%%%%%%
\subsection{Intersecting $NS$ and $D$-Branes}
%%%%%%%%%%%%%%%%%%%%%%%%
We can use the duality transformations discussed in the last subsection to
obtain all intersecting brane solutions in type II string theory.
Lets start with  the $M2\perp M2(0)$ solution with the $M2$-branes oriented
along say the $\{1,2\}$ and $\{3,4\}$ directions.
Reducing this configuration along an overall transverse
direction we obtain a $D2\perp D2(0)$ solution with the $D2$-branes
having the same orientations. If we now perform $T$-duality
in a direction parallel to one of the $D2$-branes, say the $\{2\}$
direction, we transform it into
a $D1$-brane in the $\{1\}$ direction 
and the other $D2$-brane into a $D3$-brane with orientation 
$\{2,3,4\}$. The final configuration
is thus $D1\perp D3(0)$. One can continue $T$-dualising in all 
possible ways and one
generates the following list of intersecting $D$-branes \cite{gkt,bbj}:
\bea
IIA:\quad &&2\perp 2 (0);\quad 4\perp 4 (2);\quad 6\perp 6(4);
\quad 2\perp 4 (1);
\quad 4\perp 6(3);\nonu\\
&&6\perp 8 (5);
\quad 
0\perp 4 (0);\quad 2\perp 6 (2);\quad 4\perp 8 (4);\nonu\\
IIB:\quad &&3\perp 3 (1);\quad5\perp 5 (3);\quad 7\perp 7 (5);
\quad 1\perp 3 (0);\quad 3\perp 5 (2);\nonu\\
&&5\perp 7 (4);
\quad 1\perp 5 (1);
\quad  3\perp 7 (3);
\quad  5\perp 9 (5);
\label{eq:ndfourlist}
\eea
These solutions all preserve 1/4 of the supersymmetry and can be
directly constructed using the analogue of the
harmonic function rule we used for
$M$-branes. The harmonic functions for each brane depends on the overall
transverse coordinates.
Note that for the cases where the branes overlap in a 5-brane
the overall transverse directions have shrunk to a point and the 
derived solution is
just Minkowski space. Their could however, be more general solutions
for these cases since, for example, the case $5\perp 9(5)$
corresponds to a type I $D5$-brane which does correspond to 
a classical solution.

There is an elegant way of
characterising these $D$-brane configurations in perturbation theory. 
Consider open strings with one end on each of the two intersecting
branes. The string coordinates can either be NN, DD or ND
depending on whether the coordinate has Neumann (N) or Dirichlet (D)
boundary conditions at each end. 
The number of coordinates with mixed ND boundary conditions is  
equal to the number of relative transverse directions and is 
four in all of the above 
cases. One can also show in perturbation theory that these
configurations preserve 1/4 of the supersymmetry \cite{polch}.

If we also act with $S$-duality in the type 
IIB theory we can generate solutions containing
$NS$-branes. Further acting with $T$-duality gives:
\bea
&&Dp\perp NS1 (0),\qquad\qquad 0\le p\le 8;\nonu\\
&&Dp\perp NS5 (p-1),\qquad 1\le p\le 6;\nonu\\
&& NS1\perp NS5 (1);\qquad NS5\perp NS5(3);\nonu\\
&&NS1+W;\quad NS5+W;\quad Dp+W, \quad 1\le p \le 9,
\label{eq:nslistnd4}
\eea
where the configurations in the last line correspond to pp-waves which travel
in one direction tangent to the brane and the solutions with $NS$-branes
only are valid in both IIA and IIB. Note that we have not included
Taub-NUT configurations. We also have not included ``non-marginal"
configurations that are
obtainable by employing more general $SL(2,\bZ)$ transformations.
The $M2\perp M5(1)$ and the
$M5\perp M5(3)$ solutions can both be obtained from
the above IIA solutions, as claimed earlier. 
For example, one can uplift
$NS1\perp NS5(1)$ and $D4\perp D4(2)$, respectively.

These configurations can be broadly classified into three categories:
self-intersections, branes ending on branes and branes within branes.
Let us make some general comments on each of these.
The $D$-brane self intersections 
in (\ref{eq:ndfourlist}), $Dp\perp Dp(p-2)$, and the $NS5$ self intersection
for IIA and IIB in (\ref{eq:nslistnd4}), $NS5\perp NS5(3)$,
all satisfy the $(p-2)$ self-intersection
rule that we described earlier. The second category
is where branes can end on branes, $p\perp q(p-1)$. Although the solutions
are too general to directly describe this setup we expect that this is
the physical situation they are naturally associated with in the same
way that we explained for the $M2\perp M5(1)$ configuration (\ref{eq:m2m5}).
The best understood example of branes ending on branes
is the case $NS1\perp Dp(0)$ which corresponds to
a fundamental string ending on a $Dp$-brane. Note that the end of the
fundamental string appears as either an electric or magnetic point source
in the $D$-brane world-volume. All other cases 
of the form $p\perp q(p-1)$ can be obtained by $S$- and $T$-
duality on these configurations (which thus supports the brane ending on
brane interpretation). For example $S$-duality immediately
gives the $D1\perp D3(0)$ and $D1\perp NS5(0)$ 
configurations in the
IIB theory. It is perhaps worth highlighting some other cases: 
the $D3\perp NS5(2)$ case in the IIB theory
was used in \cite{HW} to study three
dimensional gauge theories on the 2+1-dimensional intersection.
The case of a $D4$-brane ending on a $NS5$-brane, 
$NS5\perp D4(3)$ in IIA was used to study four-dimensional gauge
theories \cite{kutasov}. It is interesting that this 
can be lifted to $M$-theory as a single $M5$-brane \cite{wittenmth}.
The third category of configurations are the branes inside of 
branes which have the form
$p\perp q(p)$. These correspond to brane soliton solutions inside the
world-volume theory. As an example consider $D0\perp D4(0)$. If we consider
$N$ parallel $D4$-branes then we should consider a $U(N)$ super-Yang-Mills
theory on the $4+1$ dimensional world-volume \cite{witworldvol}. 
Four dimensional
euclidean $U(N)$ instantons correspond to static solitons in the
world-volume theory which can also be interpreted as $D0$-branes
\cite{douglas}. 

We now comment on some different configurations of branes
that give rise to $D$=4 and $D$=5 black holes.
Start with the 
$M2\perp M5(1)$ configuration with a pp-wave along the
intersection (\ref{eq:firstm2m5wave}) which gives a $D$=5 black
hole upon dimensional reduction. One can now 
perform the following steps:
\bea
\matrix{
M5:&1&2&3&4&5& & & & & \cr
M2:&1& & & & & & & & &10 \cr 
\ {}W:&1& & & & & & & & & \cr}\quad&\stackrel{R10}{\longrightarrow}&\quad
\matrix{
NS5:&1&2&3&4&5\cr
NS1:&1& & & & \cr
\ {} \ {}W:  &1& & & & \cr}\quad\stackrel{T1}{\longrightarrow}
\nonu\\
\nonu\\[20pt]
\matrix{
NS5:&1&2&3&4&5& & & & & \cr
\ {} \ {}W:&1& & & & & & & &  &\cr
NS1:&1& & & & & & & & & & \cr}\quad&\stackrel{S}{\longrightarrow}&\quad
\matrix{
D5:&1&2&3&4&5\cr
\ {}W: &1& & & & \cr
D1:&1& & & & \cr} 
\eea
where we have dimensionally reduced on the 10 direction to get a IIA solution, 
$T$-dualised
on the 1 direction to get a IIB solution and then performed $S$-duality.
The resulting IIB configuration, $D5\perp D5(1)$ plus a pp-wave \cite{calmald},
is the case that has been most studied in black hole entropy 
studies. We noted that the $M2\perp M2\perp M2$
solution (\ref{eq:twotwotwo}) 
can also give a $D$=5 black hole. It can be  
related to the above configuration by dimensional reduction and duality:
\bea
\matrix{
M2: &1& & & & & & & & &10 \cr
M2: & &2&3& & & & & & &  \cr 
M2: & & & &4&5& & & & & \cr}\quad&\stackrel{R10}{\longrightarrow}&\quad
\matrix{
N1:&1& & & & \cr
D2:& &2&3& & \cr 
D2:& & & &4&5\cr}\quad\stackrel{T145}{\longrightarrow}\nonu\\
\nonu\\[10pt]
\matrix{
\ {}W:  &1& & & & & & & & &\cr
D5: &1&2&3&4&5& & & & &\cr 
D1: &1& & & & & & & & &\cr}.\ {}\ {}\quad& \ {}&
\eea

We mentioned two ways in which $D$=4 black holes can be obtained from
intersecting $M$-branes: $M5\perp M5\perp M5$ with momentum flowing
along a common string direction (\ref{eq:fivefivefivewave}), 
and $M5\perp M5\perp M2\perp M2$ (\ref{eq:twotwofivefive}).
Both of these can be related to a very symmetrical configuration
of four $D3$-branes \cite{klebtseyt,balasdfour}. 
The second case works as follows:
\bea
\matrix{
M5:&1&2&3&4&5& & & & & \cr
M5:&1&2&3& & &6& & & &10 \cr 
M2:& & & &4& &6& & & &  \cr 
M2:& & & & &5& & & & &10 \cr}\quad&\stackrel{R10}{\longrightarrow}&\quad
\matrix{
NS5:&1&2&3&4&5& \cr
\ {}D4:&1&2&3& & &6\cr 
\ {}D2:& & & &4& &6\cr 
NS1:& & & & &5& \cr}\quad\stackrel{ST1}{\longrightarrow}\nonu\\
\nonu\\[10pt]
\matrix{
D5:&1&2&3&4&5& & & & &\ {} \ {} \cr
D3:& &2&3& & &6& & & &\ {} \ {} \cr 
D3:&1& & &4& &6& & & &\ {} \ {} \cr 
D1:& & & & &5& & & & &\ {} \ {} \cr}\quad&\stackrel{T34}{\longrightarrow}&\quad
\matrix{
\ {} \ {}D3: &1&2& & &5& \cr
\ {} \ {}D3: & &2& &4& &6\cr 
\ {} \ {}D3: &1& &3& & &6\cr 
\ {} \ {}D3: & & &3&4&5& \cr}.
\eea

We now turn to the overlapping brane solutions that can be generated
from the $M5\perp M5(1)$ overlap (\ref{eq:m5m5over}). 
Reducing on the common string direction
we get $D4\perp D4(0)$ and $T$-duality generates the list
of overlapping $D$-branes:
\bea
IIA:\quad &&0\perp 8 (0);\quad 2\perp 6 (0);\quad 2\perp 8 (1);
\quad 4\perp 4 (0);
\quad 4\perp 6 (1);\nonu\\
IIB:\quad &&1\perp 7 (0);\quad 1\perp 9 (1);
\quad  3\perp 5 (0);\quad  3\perp 7 (1);\quad  5\perp 5 (1).
\label{eq:ndeight}
\eea
These solutions all break 1/4 of the supersymmetry and
can also be directly constructed using the harmonic
function rule but taking into account that the harmonic functions
depend on the relative transverse coordinates not the overall transverse
coordinates.
At the level of string perturbation theory, these configurations
correspond to $D$-branes
that have eight string coordinates with mixed ND boundary
conditions. 
Employing $S$-duality we obtain the following configurations with $NS$ branes:
\bea
&&NS5\perp Dp(p-3)\qquad 3\le p\le 8;\nonu\\
&&NS5\perp NS5(1).
\label{eq:ndeightnsand}
\eea

Recall that an extra $M2$-brane can be added to 
the $M5\perp M5(1)$ solution 
without breaking any more supersymmetry and the resulting configuration can
be interpreted as an $M2$-brane stretched between the two $M5$-branes.
After dimensional reduction of this solution and performing dualities
we find analogous generalisations of the above solutions.
For the $D$-brane intersections in (\ref{eq:ndeight})
that intersect in a point, we find
that we can add a fundamental string without breaking any more 
supersymmetry. This can then be interpreted as a fundamental string being
stretched between the two $D$-branes. If the $D$-branes intersect in a
string then we find that we can add a pp-wave along this string
intersection without breaking any more supersymmetry. One interesting
example of this is the $1\perp 9(1)$ case. Since the IIB $D9$-brane
leads to the type I theory, our interpretation translates into
the fact
that a type I $D$-string can carry momentum without breaking any more
supersymmetry. This $D$-string can be interpreted as a heterotic string soliton
\cite{witpol} and the properties of the heterotic
string with momentum were studied in detail in \cite{dghw,calmaldpeet}.
For the cases with $NS$- and $D$-branes 
(\ref{eq:ndeightnsand}) we find that we can add 
a $D(p-2)$-brane that can be thought of as being stretched between the
$NS5$-brane and the $Dp$-brane that ends on each in a $D(p-3)$ brane,
cases that were considered above in (\ref{eq:nslistnd4}). 
One example of this is a $D3$-brane
stretched between a $NS5$-brane and a $D5$-brane, intersecting each on
a two-brane. This setup was considered in \cite{HW}. The $NS5\perp NS5(1)$
case is considered below.

In the next section we will be considering some of these configurations
but generalised so that the branes intersect at angles. In preparation for
this let us be a little more explicit about some cases. Start with
the $NS5\perp D5(2)$ configuration of IIB with the extra $D3$-brane
which preserves 1/4 of the supersymmetry and
perform $T$-duality in one of the common intersection directions:
\beq
\matrix{
N5:&1&2&3&4&5& & & &  \cr
D5:&1&2& & & & &7&8&9 \cr 
D3:&1&2& & & &6& & &  \cr}
\quad\stackrel{T2}{\longrightarrow}\quad
\matrix{
N5:&1&2&3&4&5& & & &  \cr
D4:&1& & & & & &7&8&9 \cr 
D2:&1& & & & &6& & &  \cr}
\label{eq:beginseq}
\eeq
We can now uplift this IIA solution to give the $M$-theory solution
(\ref{eq:only}):
\beq
\matrix{
M5:&1&2&3&4&5& & & & & \cr
M5:&1& & & & & &7&8&9&10 \cr 
M2:&1& & & & &6& & & & \cr}.
\label{eq:snowy}
\eeq
Reducing on the $6$ direction and then relabeling the $10$ direction as
the $6$ direction we get the IIA solution
\beq
\matrix{
NS5:&1&2&3&4&5& & & &  \cr
NS5:&1& & & & &6&7&8&9 \cr 
NS1:&1& & & & & & & & \cr}.
\label{eq:fivefivefund}
\eeq
Performing $T$-duality on the $1$ direction we get the IIB configuration
\beq
\matrix{
NS5:&1&2&3&4&5& & & &  \cr
NS5:&1& & & & &6&7&8&9 \cr 
\ {} \ {}W:&1& & & & & & & & \cr}.
\label{eq:fivefivewave}
\eeq
It is interesting to note that in the IIA theory we can add a
fundamental string to the $NS5\perp NS5(1)$ configuration without
breaking any more supersymmetry, while in the IIB theory we can add 
a pp-wave. Since the IIA and IIB theories have the same $NS$-fields
the configuration (\ref{eq:fivefivefund}) does give a solution
of the IIB theory, but it breaks 1/8 of the supersymmetry not 1/4.
Finally carrying out $S$-duality on (\ref{eq:fivefivewave}) we get
\beq
\matrix{
D5:&1&2&3&4&5& & & &  \cr
D5:&1& & & & &6&7&8&9 \cr 
\ {}W:&1& & & & & & & & \cr}.
\label{eq:dfivedfivewave}
\eeq

%%%%%%%%%%%%%%%%%%%%%%%
\section{Branes Intersecting at Angles}
%%%%%%%%%%%%%%%%%%%%%%%%

In the configurations that we have studied
so far all of the branes have orthogonal intersections. In a perturbative
$D$-brane context it has been pointed out
that certain rotations away from orthogonality lead to
configurations that still preserve some supersymmetry \cite{BDL}. 
In this section
we will summarise some recent work on constructing classical supergravity
solutions that describe such intersections \cite{ggpt}. 
Other recent work on finding configurations with non-orthogonal 
intersections will not be discussed
\cite{bc}\cite{costacvetic}-\cite{ballarsen}.   
The solutions in \cite{ggpt} which we shall 
describe are much more complicated than the ones we have seen
so far. They have a common origin in $D$=11 using toric
\HK\ manifolds. To motivate the solutions we shall first begin
by recasting some of the orthogonal solutions in a similar language.

\subsection{Taub-NUT space and Overlapping Branes}
We begin by reviewing the construction of the $D6$-brane solution
of the type IIA theory in terms of Taub-NUT space \cite{townsix}.
Taub-NUT space is a four-dimensional \HK\ manifold. That is,
the manifold admits three covariantly constant
complex structures $J^{(m)}$ and the metric is
Kahler with respect to each. Consider the \HK\ metrics 
\bea
&&ds^2=V({\bf x})d{\bf x}\cdot d{\bf x}+V^{-1}({\bf x})(d\psi+
{\bf A}({\bf x})\cdot
d{\bf x})^2,\nonu\\
&&\nabla\times {\bf A}=\nabla V.
\eea
Choosing the harmonic function $V$ to have single centre, $V=1+m/r$,
and hence ${\bf A}= m\cos\theta d\phi$, where $(r,\theta,\phi)$ are
spherical polar coordinates on $\bE^3$, gives Taub-NUT space.
The metric appears singular at $r=0$ but this is in fact a coordinate
singularity if we choose $\psi$ to be a periodic coordinate
with period $4\pi m$. The $U(1)$ isometry corresponding to shifts in
the coordinate $\psi$ is tri-holomorphic
%The isometry group of the manifold is $U(2)$. The $U(1)$
%factor corresponding to translations in the circle direction ...
i.e., the Lie derivative of the complex structures
with respect to the $U(1)$ killing field vanishes. 
%As an aside we note that
%the three complex structures transform as a triplet with respect to
%the $SU(2)$ isometry. 
The topology of each surface with fixed $r$ is a three
sphere which is a circle bundle
over a two-sphere base with $\psi$ being the coordinate of the fibre. The
global 
topology of the manifold is $\bE^4$. From the metric we note that
as $r\to\infty$ the radius of the
circle approaches $4 \pi m$ which suggests that we can use 
Taub-NUT space in a Kaluza-Klein setting \cite{sork,grossperry}.  
Since it is \HK\
the manifold is automatically
Ricci-flat and hence will solve Einstein's equations. 
We can use this
to give an $D$=11 supergravity solution  by
adding in 6+1 Minkowski space:
\bea
ds^2&=&-dt^2+dx_1^2+\dots +dx_6^2+ds^2_{\rm TN}\nonu\\
A&=&0.
\label{eq:d6brane}
\eea
If we now reduce this solution along the $U(1)$ Killing-vector
using (\ref{eq:twoe}) 
we obtain the $D6$-brane solution with metric and dilaton as in
(\ref{eq:dbranesol})
with the
non-trivial $RR$ one-form coming from the off diagonal terms in
the metric. It is worth emphasising that while the $D6$-brane
is a singular solution in ten-dimensions, it has a non-singular
resolution in $M$-theory. If $V$ is multicentred, $V=1+\Sigma m_i/r_i$
with $r_i=|{\bf x}-{\bf x}_i|$, then we obtain
the multicentre \HK\ manifolds. They are non-singular provided that no
two centres coincide. Upon dimensional reduction they give rise
to parallel $D6$-branes.

If we relabel the Taub-NUT coordinates $({\bf x},\psi)$=$(x_7,\dots,x_{10})$,
then the solution (\ref{eq:d6brane}) has 16 Killing spinors which satisfy
the constraints
\beq
\epsilon=\hat\Gamma_{78910}\epsilon.
\eeq
This is equivalent 
to the $D6$-brane constraint (\ref{eq:susyrules}).
if we reduce on $x_{10}$.
Consider now the Taub-NUT space to lie along the $(x_3,\dots,x_6)$ directions
with $x_6$ being the coordinate on the
circle. If we now reduce along $x_{10}$
then we arrive at the IIA Taub-NUT configuration. 
Recall that if we now
$T$-dualise in the circle direction $x_{6}$ we obtain the IIB 
$NS5$-brane (delocalised in the $x_{6}$
direction transverse to the brane). This will be useful in a moment.
For completeness we note here that the supersymmetry projections for 
a IIA or IIB Taub-NUT configuration in the $(x_3,\dots,x_6)$
direction is
\beq
{\rm IIA/B}\quad {\rm Taub-NUT}:\qquad\ep_L=\hat\Gamma_{3456}\ep_L
\qquad\ep_R=\hat\Gamma_{3456}\ep_R.
\eeq

A natural generalisation of the above construction of the $D6$-brane
is to consider
an eight dimensional Ricci-flat manifold obtained as the product
of two Taub-NUTS. By adding in 2+1 dimensional Minkowski space we
get a $D$=11 supergravity solution:
\bea
ds^2&=&ds^2(\bE^{1,2})+ds^2_{\rm TN_1} +ds^2_{\rm TN_2}\nonu\\
A&=&0.
\label{eq:tntn}
\eea
Label the coordinates of the circles of the two Taub-NUT metrics
by $x_6$ and $x_{10}$, respectively.
%Before interpreting this solution, lets briefly return to (\ref{eq:d6brane}).
%If we instead reduce to $D$=10 along one
%of the 6 flat directions, $x_6$ say, we obtain a IIA Taub-NUT solution.
%If we now $T$-dualise this solution in the $\psi$
%direction, the off diagonal components of the metric get
%replaced with the $NS$ two-form and we are led to the IIB
%$NS5$-brane\footnote{To be precise the 
%$NS5$-brane is delocalised in the $\psi$
%direction transverse to the brane.}.
%Returning to (\ref{eq:tntn}) 
Reduce on the $x_{10}$
direction to get a IIA configuration
and then
$T$-dualise on the $x_6$ direction to get a type IIB solution.
Reducing the second Taub-NUT on $x_{10}$ leads to a $D6$-brane in the
$1,\dots, 6$ directions and $T$-dualising in the $x_6$ direction 
transforms it into
a $D5$-brane. On the other hand, reducing the first Taub-NUT on
$x_{10}$ gives IIA
Taub-NUT and the $T$-duality turns it into a IIB $NS5$-brane. Since
both branes share the 2+1 dimensional space we see that 
the final configuration is a $D5$-brane orthogonally overlapping 
a $NS5$-brane in a two brane, $D5\perp NS5 (2)$,
which is a solution we have already
considered. Recall that by a sequence of dualities
we can relate it to the first pair of branes in 
(\ref{eq:beginseq})-(\ref{eq:dfivedfivewave}).

\subsection{Toric \HK\ Manifolds and Branes Intersecting at Angles}

To obtain solutions corresponding to
non-orthogonally overlapping branes we 
replace Taub-NUT$\times$Taub-NUT by an eight-dimensional 
toric \hk\ manifold i.e., one that
admits a $U(1)\times U(1)$ triholomorphic isometry: 
\bea
ds^2&=&ds^2(\bE^{1,2})+ds^2_{HK}\nonu\\
A&=&0.
\label{eq:introa}
\eea
After dimensional reduction and dualities 
we shall get solutions with
branes as in (\ref{eq:beginseq})-(\ref{eq:dfivedfivewave}) 
(ignoring the last entry) that overlap 
non-orthogonally. We shall discuss the inclusion of the other brane
later. 
%For the generalisation of the $NS5\perp D5(2)$
%case we will see that there is an interesting
%correlation of the angle with the type of $(p,q)$-brane of the IIB
%theory.
One interesting aspect of these solutions is that
they all come from completely regular $D$=11 metrics.

All of these solutions will generically
preserve $3/16$ of the supersymmetry. 
The proof of this is 
essentially an application of the methods 
used previously in the context of KK
compactifications of $D$=11 supergravity (see, for example, \cite{DNP}).
We first decompose the 
32-component Majorana spinor of the $D$=11 Lorentz group 
into representations of $SL(2;\bR)\times SO(8)$:
\beq
{\bf 32} \rightarrow ({\bf 2},{\bf 8}_s) \oplus ({\bf 2},{\bf 8}_c)\ .
\label{eq:twob}
\eeq
The two different 8-component spinors of $SO(8)$ correspond to the two possible
$SO(8)$ chiralities. The unbroken supersymmetries correspond to singlets in the
decomposition of the above $SO(8)$ representations with respect to the holonomy
group ${\cal H}$ of ${\cal M}$. Consider for example, $D$=11 Minkowski space
for which ${\cal H}$ is trivial; in this case both
8-dimensional spinor representations decompose into 8 singlets, so that all
supersymmetries are preserved. The generic holonomy group for an
eight-dimensional \hk\ manifold is
$Sp(2)$, for which we have the following decomposition of the $SO(8)$ spinor
representations:
\bea
{\bf 8}_s &\rightarrow& {\bf 5} 
\oplus {\bf 1}\oplus {\bf 1}\oplus {\bf 1}\nonu\\
{\bf 8}_c &\rightarrow& {\bf 4}\oplus {\bf 4}\ .
\label{eq:twoc}
\eea
There are now a total of 
6 singlets (three $SL(2;\bR)$ doublets) instead of 32, 
so that the $D$=11 supergravity solution preserves 3/16 of the supersymmetry,
unless the holonomy happens to be a proper subgroup of $Sp(2)$ in which case
the above representations must be further decomposed. For example, the ${\bf
5}$ and ${\bf 4}$ representations of $Sp(2)$ have the decomposition
\bea
{\bf 5} &\rightarrow& ({\bf 2},{\bf 2}) \oplus ({\bf 1},{\bf 1})\nonu\\
{\bf 4} &\rightarrow& ({\bf 2},{\bf 1}) \oplus ({\bf 1},{\bf 2}) 
\label{eq:twod}
\eea
into representations of $Sp(1)\times Sp(1)$. We see in this case that there are
two more singlets (one $SL(2;\bR)$ doublet), from which it follows that the
solution preserves 1/4 of the supersymmetry whenever the holonomy is
$Sp(1)\times Sp(1)$. Since this is the holomony group
for Taub-NUT$\times$Taub-NUT space, we recover our previous result.

\subsection{Toric \hk\ manifolds}
To proceed we need to be more concrete about the properties of
eight dimensional toric \hk\ manifolds. 
The most general metric has the
local form
\beq
ds^2 = U_{ij}\, d{\bf x}^i\cdot d{\bf x}^j + 
U^{ij}(d\varphi_i + A_i)(d\varphi_j + A_j),
\label{eq:onec}
\eeq
where $U_{ij}$ are the entries of a positive definite symmetric $2\times 2$
matrix function $U$ of the $2$ sets of coordinates ${\bf x}^i = \{x_r^i\, ;
r=1,2,3\}$ on each of $2$ copies of $\bE^3$, and $U^{ij}$ are the entries of
$U^{-1}$. The two one-forms $A_i$ have the form
$A_i = d{\bf x}^j\cdot \bfomeg_{ji}$
where $\bfomeg$ is a triplet of $2\times 2$ matrix functions
of coordinates on $\bE^6$ and are 
determined by the 
matrix $U_{ij}$. Specifically, the two-forms $F_i=dA_i$ with components
\beq
F_{jk}^{rs}{}_i = \partial^r_j\omega^s_{ki} - \partial^s_k\omega^r_{ji}\ ,
\eeq
must satisfy
\beq
F_{jk}^{rs}{}_i = \varepsilon^{rst} \partial^t_jU_{ki},
\eeq
where we have introduced the notation
\beq
{\partial\over \partial x_r^i} =\partial^r_i\ .
\eeq
Note that $dF_i=0$ implies 
\beq
{\bf \partial}_i\cdot {\bf \partial}_j\, U =0\qquad (i,j=1,2)\ .
\eeq

The simplest \hk\ manifold, which may be considered to
represent the `vacuum', is constant $U$ which implies $A_i=0$. 
We shall denote this constant
`vacuum matrix' by $U^{(\infty)}$.
For our applications we may restrict $U^{(\infty)}$
to be such that
\beq
\det U^{(\infty)} =1.
\label{eq:newa}
\eeq
Regular non-vacuum \hk\ metrics can be found by superposing this with
some 
linear combination of
matrices of the form
\beq
U_{ij}[\{ p\},{\bf a}] = {p_ip_j\over 2|\sum_k \, p_k{\bf x}^k\, - {\bf
a}|},
\label{eq:onelc}
\eeq
where the `p-vector' $\{p_1,p_2\}$ is an ordered set of coprime integers
and 
${\bf a}$ is
an arbitrary 3-vector. Any matrix of this form may be associated with a
3-plane in $\bE^{6}$, specified by the 3-vector equation
\beq
p_1 {\bf x}^1+p_2 {\bf x}^2 = {\bf a}.
\label{eq:oneld}
\eeq
If we have two p-vectors the angle between the two 
3-planes can be determined and is given by:
\beq
\cos\theta = {p\cdot p'\over \sqrt{p^2 p'^2}},
\label{eq:onepa}
\eeq
with inner product
\beq
p\cdot q = (U^{(\infty)})^{ij}p_iq_j.
\label{eq:onepb}
\eeq

The general non-singular metric may now be found by linear
superposition. For a given p-vector we may superpose any finite number
$N(\{p\})$ of solutions with various distinct 3-vectors $\{{\bf a}_m(\{p\});\
m=1,\dots,N\}$. We may then superpose any finite number of such solutions.
This construction yields a solution of the \hk\ conditions of the form
\beq
U_{ij} = U^{(\infty)}_{ij} + \sum_{\{p\}}\sum_{m=1}^{N(\{p\})}
U_{ij}[\{ p\},{\bf a}_m(\{p\})].
\label{eq:onele}
\eeq
Since each term in the sum is associated with a $3$-plane in $\bE^{6}$, any
given solution is specified by the angles and distances between some finite
number of mutually intersecting $3$-planes \cite{dancer}. 
It can be shown that
the resulting \hk\ $8$-metric is complete provided that no two intersection
points, and no two planes, coincide. This is the analogue of the four
dimensional multicentre metrics being singular when two centres coincide
and has been demonstrated by
means of the \hk\ quotient construction in \cite{ggpt}.

The simplest examples of these manifolds 
are found by supposing $\Delta U \equiv
U-U^{(\infty)}$ to be diagonal. For example,
\beq
U_{ij} = U^{(\infty)}_{ij} +  \delta_{ij} \, {1\over 2|{\bf x}^i|}.
\label{eq:onem}
\eeq
which is constructed from the p-vectors $(1,0)$ and $(0,1)$.
\HK\ metrics with $U$ of this form were found previously 
on the moduli space of
$2$ distinct fundamental BPS monopoles in maximally-broken rank four gauge
theories \cite{LWY} (see also \cite{manton}). 
For this reason we shall refer to them
as  `LWY metrics'. 
Whenever $\Delta U$ is diagonal we may choose 
the two one-forms $A_i$ to be one-forms on 
the $i$th Euclidean 3-space satisfying
\beq
F_i = \star dU_{ii}\qquad (i=1,2),
\label{eq:onen}
\eeq
where $\star$ is the Hodge dual on $\bE^3$. 

For the special case in which not only $\Delta U$ but also
$U^{(\infty)}$ is diagonal then $U$ is diagonal and the LWY metrics reduce to
the metric product of $2$ Taub-Nut metrics with $Sp(1)\times
Sp(1)$ holonomy. 
Note that for the LWY metrics the angle between the 3-planes, 
(\ref{eq:onepa})
reduces to
\beq
\cos\theta = -{U^{(\infty)}_{12}\over
\sqrt{U^{(\infty)}_{11}U^{(\infty)}_{22}}},
\label{eq:onepc}
\eeq
and we see that $Sp(1)\times Sp(1)$ holonomy occurs 
when the 3-planes intersect orthogonally. In general one can argue that
the holonomy of a general toric \HK\ manifold  is $Sp(2)$ and is
only a proper subgroup of $Sp(2)$ when there are only 
two 3-planes or two sets of
parallel 3-planes intersecting orthogonally, 
in which case the metric is a product
of two \HK\ 4-metrics.

\subsection{Overlapping branes from \hk\ manifolds}

Let us return to the interpretation of our $D$=11 solution (\ref{eq:introa}) 
for a general \hk\ manifold specified by a matrix $U$ as in 
the last subsection.
We follow the steps that we 
considered when we discussed Taub-NUT$\times$Taub-NUT.
We first reduce the solution along one of the $U(1)$ Killing vectors
to obtain a IIA solution that preserves 3/16 of the supersymmetry and then
$T$-dualise along the other $U(1)$ Killing vector.
Using the 
$T$-duality rules of \cite{BHO} we get a IIB solution
with Einstein metric and other fields given by
\bea
ds^2_E &=& (\det U)^{3\over4} \big[(\det U)^{-1}ds^2(\bE^{2,1}) + (\det
U)^{-1}U_{ij} d{\bf x}^i\cdot d{\bf x}^j  + dz^2\big] \nonu\\
B_{(i)} &=& A_i\wedge dz \nonu\\
\tau &=& -{U_{12}\over U_{11}} + i{\sqrt{\det U}\over U_{11}}. 
\label{eq:twok}
\eea

As the interpretation of this solution is rather subtle lets 
first consider continuing with the transformations as
in (\ref{eq:beginseq})-(\ref{eq:dfivedfivewave}):
$T$-dualising on one of the $\bE^{2,1}$ directions leads
to a IIA solution which we shall not write down. If we uplift
it to $D$=11 one obtains:
\bea
ds^2_{11} &=& (\det U)^{2\over3}\big[(\det U)^{-1} ds^2(\bE^{1,1}) + (\det
U)^{-1} U_{ij}\, d X^i \cdot dX^j  + dy^2\big]\nonu\\
F &=& F_i\wedge d\varphi^i\wedge dy,
\label{eq:twou}
\eea
where $X^i=({\bf x}^i,\phi^i)$ with $\phi^i$ the coordinates of the torus
that is (essentially) dual to the one with coordinates $\phi_i$.
We shall start by considering the case in
which $U$ is diagonal. In the simplest of these cases the 8-metric is
the metric product of two Euclidean Taub-Nut metrics, each of which is
determined by a harmonic function with a single pointlike singularity. Let 
$H_i = [1 + (2|{\bf x}^i|)^{-1}]\,$ be the two harmonic functions;
then 
\beq
U = \pmatrix{H_1({\bf x}^1) & 0\cr 0 & H_2({\bf x}^2)},
\label{eq:twop}
\eeq
and we return to the $M5\perp M5(1)$ 
solution (\ref{eq:m5m5over}).
Note that
in this derivation,
the $H_i$ are harmonic on the $i$th copy of $\bE^3$, 
rather than on the $i$th copy of
$\bE^4$ and hence each of the $M5$-branes
are delocalised in the direction between them
and in one direction tangent to the other $M5$-brane.
Next generalising to the LWY metrics (\ref{eq:onem}) 
we still interpret the
singularities in $U$ to be the locations of the two (delocalised) $M5$-branes.
Since the $M5$-branes have a string direction
in common, the configuration is determined by the relative orientation of two
4-planes in the 8-dimensional space spanned by both. Because of 
the delocalisation the angle between the two four planes is taken to be
the angle between the singular three planes 
(\ref{eq:onepc}). It can be argued that this rotation can be thought
of as an $Sp(2)$ rotation of one $M5$-brane relative to the other
in $\bE^8$ \cite{ggpt}. 
We thus conclude that the process of rotating one $M5$-brane
away from another by an $Sp(2)$ rotation preserves 3/16 supersymmetry.
In the more general case in which $\Delta U$ is
non-diagonal the solution can be interpreted as an arbitrary number of 
$M5$-branes
intersecting at angles determined by the associated p-vectors; these angles are
restricted only by the condition that the 
pairs of integers $p_i$ be coprime\footnote{Note that this condition comes 
from demanding that the \HK manifold is regular. If we just wanted to have
solutions to the supergravity equations of motion then we could allow
the $p_i$ to be arbitrary real numbers.}. 
It is an interesting open question whether these 3/16
supersymmetric solutions can be generalized to allow $U$ to depend on
all eight coordinates $\{X^{(i)},\, i=1,2\}$.

Reducing on the overall transverse coordinate we obtain
a IIA solution which we shall omit. In the simplest case 
that $U$ is diagonal as in (\ref{eq:twop}) it is the $NS5\perp NS5(1)$
solution in (\ref{eq:fivefivefund}). 
For more general $U$ there is an arbitrary number 
of $NS5$-branes intersecting at angles determined by their p-vectors
as in the $M5$-brane case. Again there is a delocalisation in
one direction tangent to each of the $NS5$-branes.
%Dimensional reducing on a
%\bea
%ds^2 &&= ds^2(\bE^{1,1}) +  U_{ij}\, dX^i\cdot dX^j\\
%B &&= A_i\wedge d\varphi^i \\
%\phi &&= {1\over2}\log \det U, 
%\label{eq:twos}
%\eea
%where
%\beq
%X^i = \{{\bf x}^i,\varphi^i\}\qquad (i=1,2).
%\label{eq:twot}
%\eeq
%and $dX\cdot dX$ is the flat metric on $\bE^4$ (but note that $U$ is still
%$T^2$ invariant so there is no dependence on the $\varphi^i$ coordinates).
%This solution represents an arbitrary number of IIA $NS$-5-branes
%intersecting on a string. 
If we now T-dualize in the common string
direction we obtain a solution involving
IIB $NS5$-branes with 
an identical interpretation.
This may be mapped to a similar configuration
involving only $D5$-branes by $S$-duality.
In this way we deduce that 
\bea
ds^2_E &=& (\det U)^{1\over4} \big[ ds^2(\bE^{1,1}) +  
U_{ij}\, dX^i\cdot dX^j\big] \nonu\\
B' &=& A_i\wedge d\varphi^i \nonu\\
\tau &=& i\sqrt{\det U} \ ,
\label{eq:dba}
\eea
is also solution of IIB supergravity preserving 3/16 supersymmetry. In the
simplest case, in which $U$ is of LWY type, this solution represents the
intersection on a string of two $D5$-branes, with one rotated relative to the
other by an $Sp(2)$ rotation 
with angle $\theta $, given by (\ref{eq:onepc}). 
We are now in a position to make contact with the work of Berkooz,
Douglas and Leigh \cite{BDL}. They considered two intersecting
Dirichlet (p+q)-branes with a common q-brane overlap in perturbation
theory. According to their
analysis, each configuration of this type is associated with an element of
$SO(2p)$ describing the rotation of one (p+q)-brane relative to the other in
the 2p-dimensional relative transverse space. 
The identity element of $SO(2p)$ corresponds to parallel branes, which preserve
1/2 the supersymmetry. Other elements correspond to rotated branes. The
only case considered explicitly in \cite{BDL} 
was an $SU(p)$ rotation, but it was
noted that the condition for unbroken supersymmetries was analogous to the
reduced holonomy condition arising in KK compactifications. The case we are
considering 
corresponds to an $Sp(2)$ rotation in $SO(8)$. The analysis of \cite{BDL} 
was generalised in \cite{ggpt} to show that this setup preserves 3/16
supersymmetry. In addition the solution (\ref{eq:dba}) shows that at least
for the $Sp(2)$ case the analogy with holonomy is exact since this 
IIB solution is dual to a
non-singular $D$=11 spacetime of $Sp(2)$ holonomy.

Let us now return to the interpretation of the IIB solution (\ref{eq:twok}).
When $U$ is diagonal
we obtain the $NS5\perp D5(2)$ solution. Since
the two fivebranes share two common directions, the singular three
planes correspond to the location of the fivebranes in the
six-dimension space. For this case the fivebranes are just delocalised in the
extra direction that separates them i.e., there is no further 
delocalisation in directions tangent to the other brane as above. 
%Let us discuss the interpretation of these solutions in the
%case that $U^{(\infty)}=1$ (comments on the more general case 
%can be found in ..).
By studying the action of $SL(2,\bZ)$ on the solution and recalling
that a IIB $(p,q)$ 5-brane can be constructed using $SL(2,\bZ)$
transformations, we come to
the following interpretation for a general \HK\ metric:
a `single 3-plane solution' of the \hk\ conditions with p-vector $(p_1,p_2)$ is
associated with a IIB superstring 5-brane with 5-brane charge vector
$(p_1,p_2)$. 
This implies that there is a direct correlation between the angle at
which any given 5-brane is rotated, relative to a $D5$-brane, and its 5-brane
charge.
An instructive
case to consider is the three 5-brane solution involving a $D5$-brane and an
$NS$-5-brane, having orthogonal overlap, and one other 5-brane. As the
orientation of the third 5-brane is changed from parallel to the $D5$-brane to
parallel to the $NS$-5-brane it changes, chameleon-like, from a $D$-brane to an
$NS$-brane.

\subsection{Intersecting branes from \hk\ manifolds}
There is a generalisation of (\ref{eq:introa})
%will correspond to adding in the third branes in
%(\ref{eq:beginseq}-\ref{eq:fivefivewave}) to the configurations 
%in which the first two overlap at angles.
called a `generalized membrane'
solution which takes the form
\bea
ds^2 &=& H^{-{2\over3}}ds^2(\bE^{2,1}) + H^{1\over3}ds^2_{HK}\nonu\\
F &=& \pm\omega_3 \wedge dH^{-1}, 
\label{eq:introb}
\eea
where $\omega_3$ is the volume form on $\bE^{2,1}$ and $H$ is a 
$T^2$-invariant\footnote{This condition on $H$ is
needed for our applications; it is not needed to solve the $D$=11 supergravity
equations.}
harmonic function on the \hk\
8-manifold. Provided the sign of the expression for the four-form $F$ in 
(\ref{eq:introb})
is chosen appropriately it can be shown that the solution with 
$F\ne0$ breaks no more
supersymmetries than the solution (\ref{eq:introa}) with $F=0$. 
%For
%\HK\ manifolds with Sp(2) holonomy it will therefore break 3/16 of
%the supersymmetry.
Point singularities of $H$ are naturally interpreted as the
positions of parallel $M2$-branes. For our purposes we require $H$ to be
independent of the two $\varphi$ coordinates, so singularities of $H$ will
correspond to $M2$-branes delocalized on $T^2$. Such functions satisfy
\beq
U^{ij}{\bf \partial}_i\cdot {\bf \partial}_j H=0.
\label{eq:threeb}
\eeq

Proceeding as before we can now convert this $D$=11
configuration into various intersecting brane configurations.
Lets first consider the case of the \HK\ 
manifold being a product of two Taub-NUT
manifolds. Recall that we first reduced on one of the Taub-NUT
circles and then $T$-dualised on the other circle and for
$H=1$ we obtained the $NS5\perp D5(2)$ configuration. For
$H\ne 1$, the reduction gives a $D2$-brane and the $T$-duality converts it 
into a $D3$-brane and we arrive at the first configuration in
(\ref{eq:beginseq}). Continuing with the various dualities we arrive
at all of the configurations in (\ref{eq:beginseq})-(\ref{eq:dfivedfivewave})
In the case of (\ref{eq:snowy}) we recover the $M$-theory
solution considered in (\ref{eq:only}). Note that substituting (\ref{eq:twop})
into (\ref{eq:threeb}) produces (\ref{eq:onenext}).

If we now consider a general toric \HK\ manifold in (\ref{eq:introb})
we obtain solutions corresponding to the configurations in 
(\ref{eq:beginseq})-(\ref{eq:dfivedfivewave}) with
the first two branes overlapping non-orthogonally and the third brane
stretched between them.

%%%%%%%%%%%%
\section{Conclusions}
%%%%%%%%%%%%%

In this paper we have reviewed various supergravity solutions corresponding
to BPS intersecting branes in $M$-theory and in type II string theory. 
We first discussed three basic intersections of two $M$-branes which
all break 1/4 of the supersymmetry: 
$M2\perp M2(0)$, $M2\perp M5(1)$ and $M5\perp M5(3)$. Upon dimensional
reduction these configurations were related to type II intersecting 
$D$-branes with
four ND string coordinates, as well as to
other configurations involving $NS$-branes. We argued that these
solutions could be interpreted in one of three ways: self
intersections of branes in $(p-2)$ dimensions, 
branes within branes, or branes ending on branes.
All of the solutions have
the property that they are delocalised along the relative transverse 
directions.
We pointed out that the harmonic function of one of these branes can be
generalised to have a dependence on the coordinates tangent to the other
brane.  It would be interesting if more general solutions of
(\ref{eq:onenext}) could be found.
More generally it would be of interest to construct fully localised 
intersecting brane solutions. 

We noted that multi intersections of $n$-branes are allowed and that
they generically break $2^{-n}$ of the supersymmetry. An interesting 
exception to this are various special triple overlaps that allow an
extra brane to be added without breaking more supersymmetry. We showed that
intersecting branes can be dimensionally reduced to give black holes with
non-zero horizon area in $D$=4 and $D$=5. Considering the intersecting
$D$ brane configurations from a perturbative point of view has had 
remarkable success at reproducing the black hole entropy from state counting. 
%Postulating $M$-brane dynamics in order to
%generate the same counting in $M$-theory
%is an interesting way of trying to learn more about
%$M$-brane dynamics \cite{klebtseyt}.

We also discussed the $M5\perp M5(1)$ overlap. 
This solution has the interesting property
that the $M5$-branes are localised inside the world-volume of
the other brane, but are delocalised in the direction that separates
them. We noted that these configurations violate the $(p-2)$
self intersection rule and that the resolution of this is the fact 
that there are more general solutions with an extra $M2$-brane
that still preserve 1/4 of
the supersymmetry. The extra $M2$-brane is interpreted as being stretched
between the two $M5$-branes. 
After dimensional reduction these are related
to $D$-brane intersections with the number of ND string coordinates
being eight.

We showed that toric \hk\ manifolds can be used to construct generalisations
of the $M5\perp M5(1)$ solution which preserve 3/16
supersymmetry where the $M5$-branes overlap non-orthogonally. 
Similar configurations can be obtained by dimensional reduction and
duality.
One interesting case is two $D5$-branes intersecting 
non-orthogonally. The two $D5$-branes
are related by an Sp(2) rotation in the eight relative transverse
directions. Since the solution is related by duality to a non-singular
$D$=11 spacetime of $Sp(2)$ holonomy it makes precise  
the analogy between the fraction of supersymmetry preserved 
by non-orthogonal $D$-branes and the standard holonomy
argument in Kaluza-Klein compactifications that was discussed in \cite{BDL}.
In view of this it would be
of interest to consider other subgroups of $SO(8)$. As pointed out in 
\cite{BDL},
the holonomy analogy would lead one to expect the existence of intersecting
$D$-brane configurations in which one $D$-brane is  rotated relative to another by
an $SU(4)$, $G_2$ or ${\rm Spin} (7)$ rotation matrix. If so, there presumably
exist corresponding solutions of IIB supergravity preserving 1/8, 1/8 and 1/16
of the supersymmetry, respectively. These IIB solutions would presumably have
M-theory duals, in which case one is led to wonder whether they could be
non-singular (and non-compact) $D$=11 spacetimes of holonomy $SU(4)$, $G_2$ or
${\rm Spin}(7)$.

By considering a generalised membrane solution involving the
toric \HK\ manifold allowed us to construct solutions
corresponding to branes overlapping non-orthogonally 
with an additional brane stretched in between them.
In the case in which a $D3$-brane intersects overlapping IIB 
5-branes, the fact
that the solution preserves 3/16 of the supersymmetry implies 
that the field theory on the 2-brane intersection has
$N$=3
supersymmetry.
When the 5-branes overlap orthogonally the field theory has
$N=4$ supersymmetry and for this case 
Hanany and Witten have shown that the brane point
of view can be used 
to determine the low-energy effective actions of 
these field theories \cite{HW}.
It would be interesting if these techniques could be adapated to
the $N=3$ case when the 5-branes overlap non-orthogonally. 

We hope to have given the impression that although much is known
about intersecting branes there is still much to be understood.

%\vfill\eject\null

\section*{Acknowledgments}
I would like to thank Fay Dowker, Gary Gibbons, David Kastor,
Juan Maldacena,
George Papadopoulos, Paul Townsend, Jennie Traschen and Arkady Tseytlin
for enjoyable
collaborations and fruitful discussions.

\end{document}